\documentclass[preprint]{iucr}              
     \journalcode{S}              
\usepackage{xcolor}
\usepackage{enumitem}
\usepackage{amsmath,mathtools,calc}
\usepackage{url}
\usepackage{ulem}
\usepackage{amssymb}
\usepackage[utf8]{inputenc}
\graphicspath{{figs/}}

\DeclareMathOperator*{\argmin}{arg\,min}
\DeclareMathOperator{\Lop}{\mathcal{L}}

\begin{document}                  
\title{Laminography as a tool for imaging large-size samples with high resolution}
\cauthor[a]{Viktor}{Nikitin}{vnikitin@anl.gov}{{address if different from \aff}}
\author[b]{Gregg}{Wildenberg}
\author[a]{Alberto}{Mittone}
\author[a]{Pavel}{Shevchenko}
\author[a]{Alex}{Deriy}
\author[a]{Francesco}{De Carlo}
\aff[a]{Advanced Photon Source, Argonne National Laboratory, 60439 Lemont, IL, \country{USA}}
\aff[b]{University of Chicago, 60637 Chicago, IL, \country{USA}}

\keyword{Laminography}
\keyword{Reconstruction}
\keyword{GPU}
\keyword{Neuroimaging}
\maketitle                        

\begin{synopsis}
We present the laminography setup and software toolbox for efficient imaging of large-size samples. Applicability of the developed technique is demonstrated with imaging the slabs through the whole mouse brain sample stained with heavy elements. \end{synopsis}

\begin{abstract}
Despite the increased brilliance of the new generation synchrotron sources, there is still a challenge with high-resolution scanning of very thick and absorbing samples, such as the whole mouse brain stained with heavy elements, and, extending further, brains of primates. 
Samples are typically cut into smaller parts, to ensure a sufficient X-ray transmission, and scanned separately. Compared to the standard tomography setup where the sample would be cut into many pillars, the laminographic geometry operates with slab-shaped sections significantly reducing the number of sample parts to be prepared, the cutting damage and data stitching problems. 
In this work, we present a laminography pipeline for imaging large samples ($\textgreater$ 1 cm) at micrometer resolution. The implementation includes a low-cost instrument setup installed at the 2-BM micro-CT beamline of the Advanced Photon Source (APS). Additionally, we present sample mounting, scanning techniques, data stitching procedures, a fast reconstruction algorithm with low computational complexity, and accelerated reconstruction on multi-GPU systems for processing large-scale datasets. The applicability of the whole laminography pipeline was demonstrated with imaging 4 sequential slabs throughout the entire mouse brain sample stained with osmium, in total generating approximately 12~TB of raw data for reconstruction.
\end{abstract}

\section{Introduction}

The computed laminography technique is an extension of regular computed tomography, which involves tilting the rotary stage along the beam direction~\cite{hasenkamp1973radiographic, gondrom1999x}. This allows for scanning planar and high X-ray absorbing samples with enhanced quality and less radiation damage. 

Laminography imaging has been demonstrated at several synchrotron facilities and lab-CT systems around the world.   
Helfen et.al implemented the technique at several beamlines of the European Synchrotron Radiation Facility~(ESRF)~\cite{helfen2005high, helfen2011implementation,helfen2013laminographic}. The method has become a routine tool at the synchrotron by showing high-quality results for many kinds of samples~\cite{xu2010synchrotron, reischig2013high,morgeneyer2014situ}. Recently, the first in-situ nano-laminography has been demonstrated with the Projection X-ray Microscope at the ESRF~\cite{hurst2023hierarchically}. Furthermore, multi-contrast computed laminography was shown at a beamline of Karlsruhe Research Accelerator~(KARA)~\cite{cheng2013multi}. The authors used a grating interferometer to enhance phase contrast of a butterfly sample.
Hoshino et. al demonstrated laminography at Spring-8 by analyzing a copper grid pattern with alphabetical letters~\cite{hoshino2011development}. At the Swiss Light Source~(SLS), the laminography geometry has been also used in nano-resolution 3D ptychographic imaging of integrated circuits~\cite{holler2019three,holler2020lamni}.
Laminography has become popular also for lab-CT sources with cone X-ray beam. Different cone-beam laminography trajectories were compared in~\cite{o2016comparing}.  In~\cite{fisher2019laminography} it is demonstrated a computed laminography implementation on a conventional industrial laboratory micro-CT scanner (Nikon XTEK), without the need for special equipment. The authors also presented methods for reducing laminography artefacts due to insufficient sampling. Another custom build X-ray CT scanner was introduced in~\cite{deyhle2020spatial}, together with a detailed guidance on the instrument calibration and optimal data acquisition. 
Furthermore, recent progress in robotic sample manipulator systems has facilitated the adjustment of laminography scanning geometry~\cite{wood2019computed}.

Several reconstruction software packages have functionality for laminography reconstruction. In the Astra Tomography Toolbox~\cite{van2015astra}, the laminography geometry can be explicitly defined, followed by running iterative reconstruction (e.g., SIRT or CGLS method) on GPUs. For an iterative method, necessary data are typically kept in GPU memory during all iterations, minimizing overhead for data copy between the CPU and GPU memory.
In such cases, the performance of reconstruction is mostly limited by GPU computation speed. Another package, called UFO~\cite{farago2022tofu}, provides a multi-threaded, GPU-enabled and distributed data processing framework for tomographic and laminographic reconstruction. Both packages, Astra and UFO, implements laminography reconstruction by direct discretization of the forward and backward projection line integrals. In this case, computational complexity is $\mathcal{O}(N^4)$ assuming that the number of projection angles and volume size in each dimension are of the order of $N$. 

Computational complexity for laminography reconstruction can be decreased to $\mathcal{O}(N^3\log N)$ using a Fourier-based method and Fast Fourier Transform (FFT), similar to the one used for regular tomography by Gridrec algorithm~\cite{Dow:99} in TomoPy~\cite{gursoy2014tomopy}, or by Fourierrec in TomocuPy~\cite{nikitin2023tomocupy}. The tomography back-projection operator can be rapidly evaluated as a combination of  the one-dimensional FFT and two-dimensional unequally-spaced inverse FFT~\cite{beylkin1998applications}. In turn, rapid evaluation of the laminography back-projection operator can be done as a combination of the two-dimensional FFT and three-dimensional unequally-spaced inverse FFT~\cite{myagotin2013efficient,voropaev2016direct}.
Computational complexity plays an important role when reconstructing data obtained by stitching several projection datasets~\cite{vescovi2018tomosaic}. For instance, performance tables in ~\cite{nikitin2023tomocupy} show that tomographic reconstruction by a method with lower complexity for 2048$^3$, 4096$^3$, 8192$^3$ volumes becomes faster than the direct discretization by factors 5, 10, 20, respectively. Current data storage allow working with stitched data of more than 16384 pixels in each dimension (more than 16~TB in single precision), therefore algorithms with lower computational complexity indeed become critical for any future tomography and laminography applications.

Nvidia GPUs have demostrated to be an essential tool in accelerating computational imaging programs. VRAM (current maximum is 80~GB for Tesla H100) is significantly smaller than computer RAM, therefore in most cases additional data splitting and transfer mechanisms have to be performed before reconstruction. These mechanisms are straightforward for regular computed tomography since each z-slice can be processed independently. But in laminography, more complex data handling procedure have to be developed since there is a dependence between slices due to the tilted geometry. Moreover, laminography slabs have typically larger sizes in two dimensions, that do not fit the detector field of view. This requires the implementation of a mosaic scanning protocol, where the slab is scanned at different positions and projection data are stitched to form a large data volume for further reconstruction. This is why it is fundamental to have fast GPU-based reconstruction with low computational complexity.  

In this work, we consider the laminography technique as a tool for optimal scanning of large-size samples. Imaging very thick and absorbing samples requires cutting these samples into parts to ensure sufficient X-ray transmission, and scanning each part independently. 
Compared to the standard tomography setup where the sample would be cut into pillars, the laminographic geometry uses slab-shaped samples, which requires significantly lower number of cutting procedures. With proper slab thickness, the X-ray propagation distance through the slab can be sufficient to obtain acceptable photon counts for different materials. 

Our contribution through this paper can be summarized as follows:  

\begin{enumerate}
\item \textit{Description of the synchrotron laminography setup.} We report how laminography is implemented at the micro-CT beamline 2-BM of the APS. The simplicity and adaptability of this setup make it an ideal choice for implementation at other synchrotron beamlines. 
\item \text{GPU-based laminography reconstruction.} We implement laminography reconstruction using the Fourier-based method ($\mathcal{O}(N^3\log N)$) on GPU with efficient asynchronous data processing by chunks where CPU-GPU data transfers and GPU processing are timely overlapped almost fully hiding time for data transfers. 
\item \textit{Integrating laminography with TomocuPy.} We add the optimized laminography reconstruction in the TomocuPy package\footnote{\url{https://tomocupy.readthedocs.io} -- \textit{in the 'develop' github branch during the paper review}}. 
TomocuPy provides an easy-to-use command-line interface for GPU-based reconstruction. Besides reconstruction of full volumes, it also provides functionality for adjusting the rotation axis and laminography tilt angle.
\item \textit{Iterative reconstruction with regularization.} Fast implementation of forward and adjoint laminography operators can be used for constructing iterative schemes with regularization. We provide, as an independent package, the implementation of reconstruction with Total Variation regularization for suppressing laminography artefacts due to insufficient sampling.
\item \textit{Scanning and reconstruction of large samples with laminography.}
Laminography simplifies the sample preparation process by requiring significantly less cutting compared to regular tomography, although more complex data stitching and reconstruction techniques are required in case of large samples. We will describe the whole pipeline employed with large samples using as example the imaging of 4 sequential mouse brain sections with $\mu$m resolution. The brain section datasets are made available in TomoBank~\cite{DeCarlo_2018}~\footnote{https://tomobank.readthedocs.io/ under Laminography/Brain}.
\end{enumerate}

\section{Strategy for scanning large-size samples}
We will start with a brief discussion on strategies for scanning large-size samples with high resolution. As an example let us consider micro-resolution imaging of an adult mouse brain sample. Typical mouse brain has sizes 12~mm (anterior to posterior)~$\times$~11~mm~$\times$~8~mm (dorsal to ventral). To increase X-ray absorption contrast in the projections, the brain is often stained with heavy metals (Osmium, Lead)~\cite{hua2015large} especially if further investigations (i.e. electron microscopy) are required. Nevertheless, high concentration of heavy materials leads to a strong X-ray beam absorption by the sample. In this case, a possible solution is to use higher energy X-ray beam, which however also significantly affects the achievable spatial resolution and sensitivity to variations in attenuation coefficient for large samples~\cite{grodzins1983optimum,flannery1987three}. 
Based on our calculation, scanning the stained mouse brain may require energy of more than 60 keV to get acceptable photon counts on the detector. For such energy, the X-ray flux of bending magnet beamlines of APS is approximately 10 times lower than the one at 20-25 keV (optimal energies). Moreover, all X-ray imaging components (scintillators, monochromator and mirror) become less efficient. The situation becomes even more difficult for larger samples, as, for example, in the cases of primate brains.

Cutting large samples into smaller sections and scanning each section independently, followed by reconstruction and stitching procedures, is the only way to handle such samples. Minimizing the number of sample cuts is of great interest because each cut damages the sample structure and causes discontinuities among sections, thereby compromising the quality of segmentation and structures tracing. This can be of critical relevance as for example in axons tracing to study the anatomy of neuronal pathways in normal and pathological states~\cite{mizutani2016three,foxley2021multi,wildenberg2023pipeline}.

\begin{figure}
    \centering
    \includegraphics[width=0.8\textwidth]{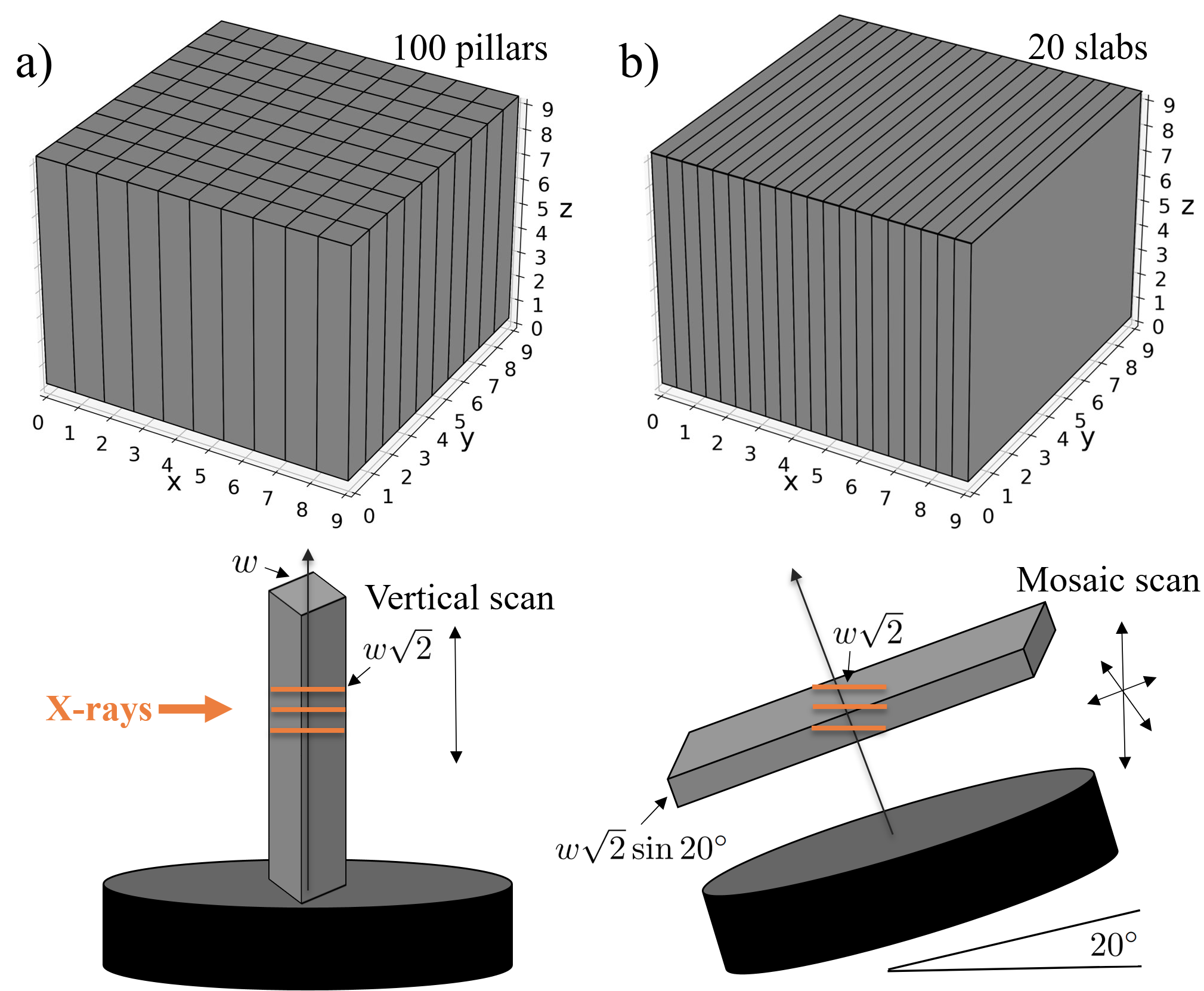}
    \caption{Schemes for scanning large samples by cutting them into parts with sufficient X-ray transmission: a) tomography geometry with pillar-shaped parts, b) laminography geometry with slab-shaped parts. }
    \label{fig:pillars_slabs}
\end{figure}

For regular tomography, samples are cut into pillars sufficiently thin to fit the experimental requirements. The pillars are then scanned at different vertical positions with overlaps. Reconstructions from the data acquired at these positions are then stitched to form the whole volume, see Figure~\ref{fig:pillars_slabs}a. Alternatively, the sample can be cut into slabs and scanned in the laminography geometry~(Figure~\ref{fig:pillars_slabs}b). The whole slab is scanned at different horizontal and vertical positions with overlaps (mosaic scanning mode,~\cite{Du:18}). The acquired data are then stitched together to form a big data volume for further laminography reconstruction. Following the sketch in the bottom part of Figure~\ref{fig:pillars_slabs}, the slab thickness is chosen as $w\sqrt{2}\sin\varphi$ where $w$ is the pillar width guaranteeing sufficient X-ray transmission along the maximum thickness $w\sqrt{2}$, and $\varphi$ denotes the laminography angle ($\varphi=20^\circ$ in the figure). The optimal thickness can be found experimentally by analyzing photon statistics on the detector and using the principles from~\cite{grodzins1983optimum,flannery1987three}. For the example in Figure~\ref{fig:pillars_slabs} with $\varphi=20^\circ$ laminography angle, the total number of sections to be cut is approximately 5 times lower in the laminography case compared with tomography (100 vs 20); for $\varphi=30^\circ$, the number of slabs is 7 times lower. Similar estimations for larger sample sizes reveal even greater improvements in the effectiveness of the laminography method, establishing it as a crucial X-ray imaging technique significantly minimizing sample damage.

\section{Laminography setup at micro-CT beamline 2-BM}

\begin{figure}
    \centering
    \includegraphics[width=1\textwidth]{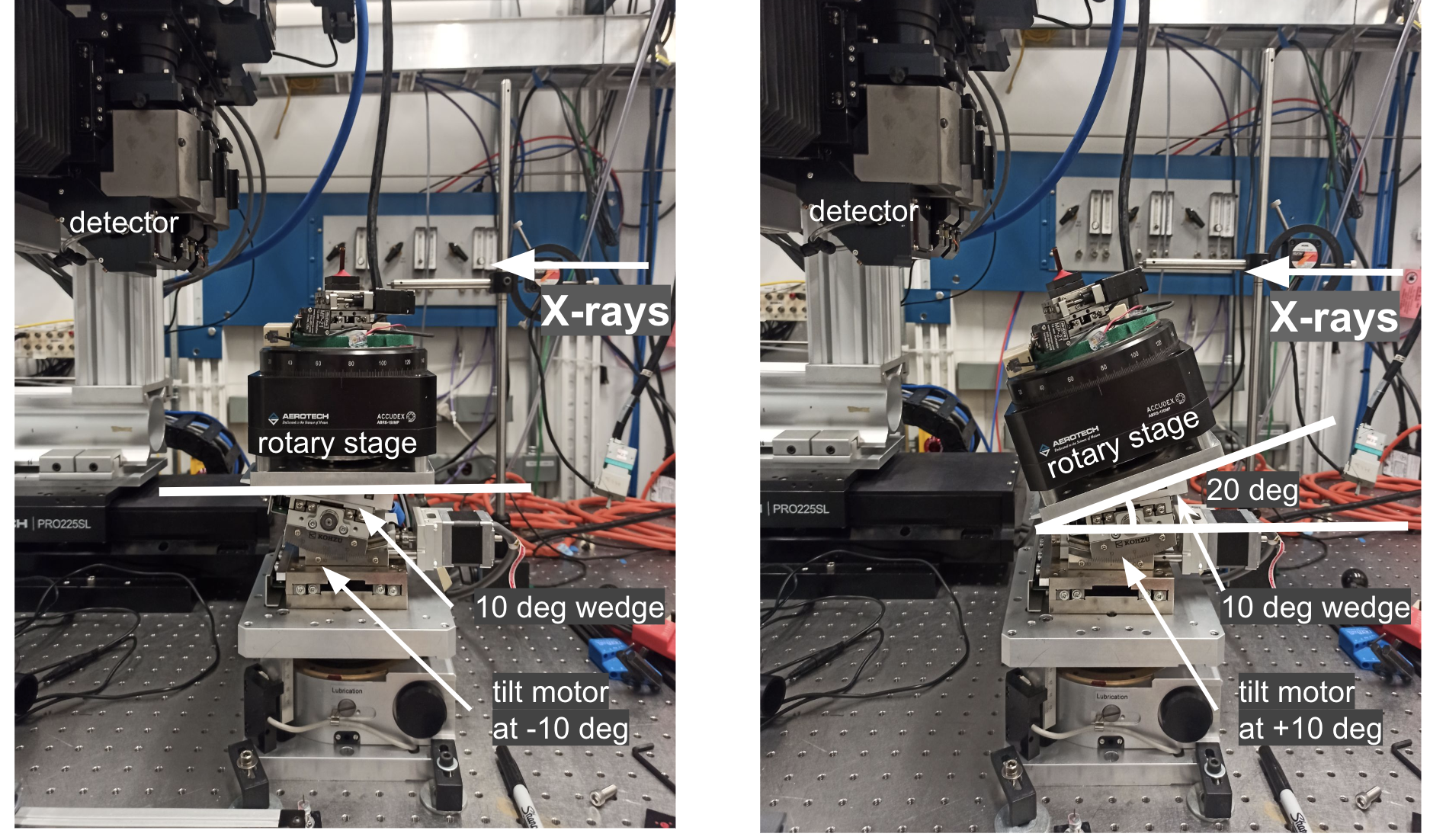}
    \caption{Sample stack with automatized switching between tomography (left) and laminography (right) geometries for conducting synchrotron experiments at sector 2-BM of the Advanced Photon source.}
    \label{fig:setup2bm}
\end{figure}

Most micro-CT beamlines at synchrotrons worldwide have similar setups for conducting experiments. They include a detection system, rotary stage, and vertical/linear stages for alignment. Besides, often the setup includes tilt motors needed for adjusting the pitch (along the beam) and roll (orthogonal to the beam) angles to properly align the rotary stage for tomographic acquisition.  
The sample is placed on the top of the rotary stage, where a micro-positioning system is normally present for selecting the region of interest for scanning.  An example of the sample stack motors implemented at beamline 2-BM of the APS is shown in Figure~\ref{fig:setup2bm}.

In order to implement the laminography geometry for scanning, one needs to tilt the rotary stage at a significant angle with respect to the beam direction. Generally, the tilt motor located under the stage allows motion only for a limited range of angles because it is used to align the rotary stage parallel with respect to the beam. The direction of the beam is adjusted for instance after switching between the pink beam and monochromatic beam, or when changing the energy on the monochromator. This alignment correction is typically less than a couple of degrees.  Most popular, compact and inexpensive tilt stages available on the market allow for travelling $\pm$10 or $\pm$15$^\circ$. At 2-BM we use a Kohzu SA07A-R2L stage with $\pm$10$^\circ$ travel range.   
Therefore, to achieve the $20^\circ$ tilt angle required by laminography, we machined a $10^\circ$  wedge and placed it under the rotary stage, see Figure~\ref{fig:setup2bm}, right. 
The wedge does not need to have precise angular specification because the fine alignment can be achieved by adjusting the tilt/roll motors under the rotary stage and analyzing the X-ray projections of a simple object like a tungsten pin. For instance, if the wedge is not perfectly flat then it is possible to compensate the miss-alignment using the motor that tilts the rotary stage orthogonal to the beam (roll alignment). 
Additional inaccuracies can be identified either through the use of a tungsten pin or by analyzing the reconstructions.

Quick switching between tomography and laminography geometry during the beamline operation is important because, first, it reduces data acquisition delays during the switch over, and second, it allows for more flexible data collection allowing to use the best geometry for the sample at hand and ultimately delivering higher quality 3D sample representation. In the current setup at 2-BM, the laminography geometry with $20^\circ$ tilt angle is achieved by placing the stage to $+10^\circ$, while positioning the stage angle to $-10^\circ$  gives us $0^\circ$ tilt against the beam, i.e. the regular tomography geometry. It is worth to mention, that this quick switch between geometries also makes the alignment procedures easier. For instance, procedures such as the rotation axis alignment and adjusting the roll stage angle (tilt orthogonal to the beam) can be first done in the tomography geometry and then reused in laminography. The misalignment issues can also be resolved in the laminographic reconstruction process, as will be shown in the next section.

\begin{figure}
    \centering
    \includegraphics[width=0.6\textwidth]{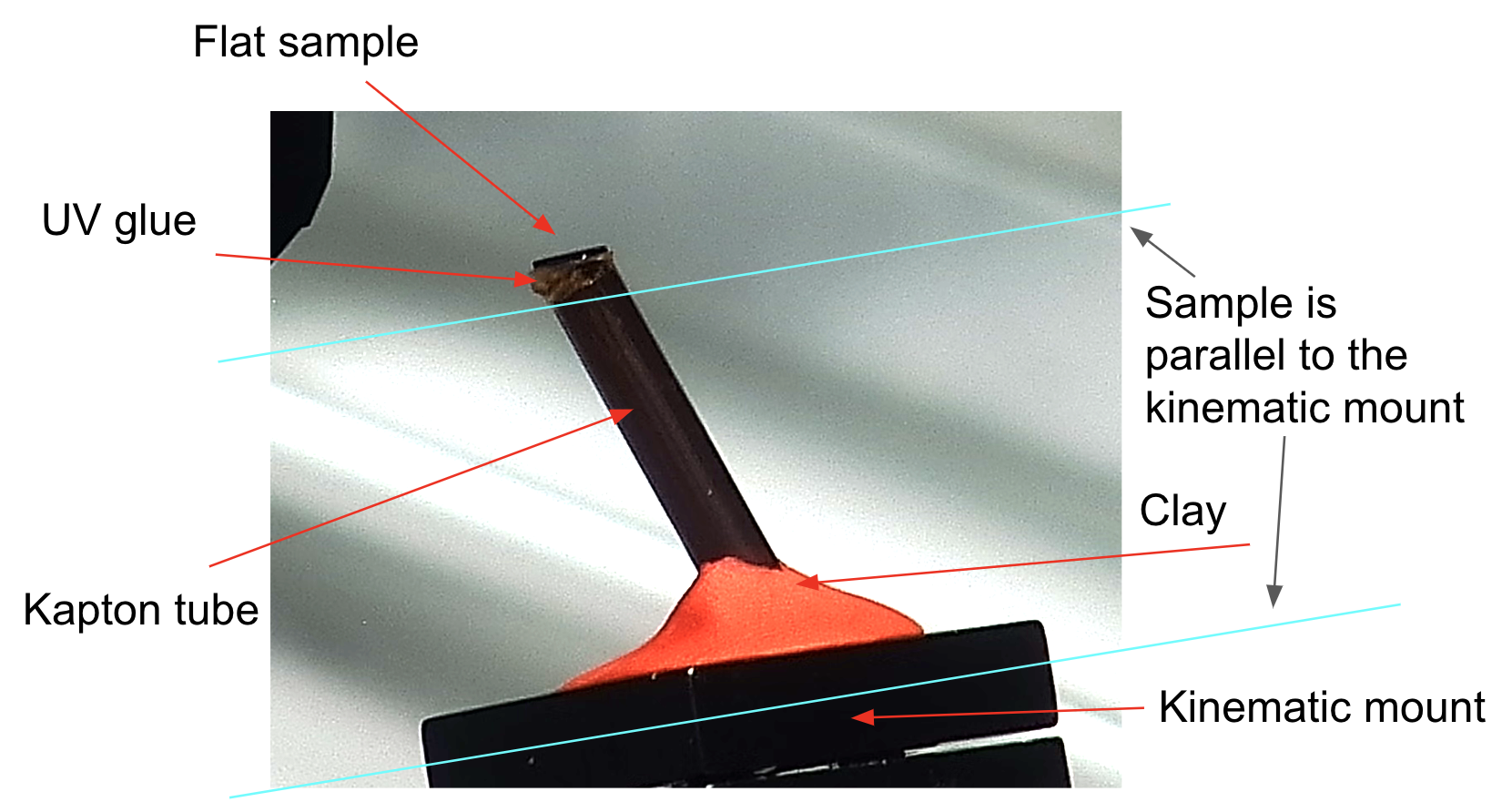}
    \caption{An example of sample mounting for laminography data acquisition.}
    \label{fig:samplemount}
\end{figure}

Another important aspect of our laminography implementation is the sample mounting procedure. Regular mounting strategies, such as gluing to a pin or fixing in a holder, are not applicable because the pin or holder will block the beam and will not allow to capture informative projections for many regions. Instead, we propose using Kapton\textsuperscript{\textregistered} tubes that are semi-transparent to X-rays, see Figure~\ref{fig:samplemount}. Kapton\textsuperscript{\textregistered} tubes with ~200~$\mu$m wall thickness are stiff enough to keep the sample stable during rotation, and do not significantly attenuate the X-ray beam reaching the detector when working with hard X-rays ($>$~10~keV). The diameter of the tube, as well as the wall thickness, can be chosen based on the sample shape and weight.
Flat samples are glued to one side of the tube with epoxy or with a UV glue supplied for instance by Bondic\textsuperscript{\textregistered} company. In our experience, the UV glue is less radiation sensitive and more transparent to X-ray than epoxy. Moreover, the glue is much easier to use since it does not have any timing requirements for mounting.
While the sample is glued to one side of the Kapton\textsuperscript{\textregistered} tube, another side of the tube is attached to a kinematic mount with clay. Alternatively, one could also use the UV glue for this. For more efficient imaging, flat samples should be mounted parallel to the kinematic mount, otherwise the X-ray propagation distance through the sample may be significantly increased for some angles, resulting in potential beam blockage.

\section{Laminography reconstruction}
In this section, we will formulate the laminography reconstruction problem in terms of operators and discuss methods for fast evaluation of these operators. 

The forward laminography operator, or laminographic projection, maps a 3D object attenuation function $\mu(x_1,x_2,x_3)$ to data $d(\theta,u,v)$, where $u,v$ are detector coordinates, and $\theta$ is the rotation angle. In this work we define the laminography tilt angle $\varphi$ as the angle between the rotation axis and beam direction (horizontal). Note that in some literature, this angle is measured between the rotation axis and the axis orthogonal to the beam (vertical). For this, variable $\varphi$ should be changed to $90-\varphi$ in all further formulas. 
We defined the laminography projection as follows,

\begin{equation}\label{eq:fwd}
\begin{aligned}    
    d(\theta,u,v) = &\Lop_{\varphi} \mu(\theta,u,v)= \int_{\mathbb{R}^3}\!u(x_1,x_2,x_3)\tilde{\delta}(\varphi,\theta,u,v)\text{d}x_1\text{d}x_2\text{d}x_3,
\end{aligned}
\end{equation}
where $\tilde{\delta}(\varphi,\theta,u,v)$ is a multiplication of two delta function defining line directions,
\begin{equation}
\tilde{\delta}(\varphi,\theta,u,v) = \delta(x_1\cos\theta\!+\!x_2\sin\theta\!-\!u)\delta(x_1\sin\theta\sin\varphi\!-\!x_2\cos\theta\sin\varphi\!+\!x_3\cos\varphi - v).
\end{equation}

The measured signal on the detector is linked to the intensity transmitted through the sample (following the Beer-Lambert  law) and also includes contribution from the dark field $d_d(u,v)$ (image on the detector when the beam is off) and flat field $d_f(u,v)$ (image on the detector when the beam is on and the sample is out), $\tilde{d}(u,v) = e^{-d(u,v)}(d_f(u,v)-d_d(u,v))+d_d(u,v)$. Therefore, before solving the inverse problem for \eqref{eq:fwd} the dark-flat field correction and taking the negative logarithm procedures are applied to the raw detector data. Note that for $\varphi=0$, the integral in \eqref{eq:fwd} becomes a general Radon transform used in tomography.

The inversion formula is given by means of filtered backprojection (FBP),
\begin{equation}\label{eq:recon}
    \mu(x_1,x_2,x_3) = \mathcal{W}\Lop_\varphi^*d(x_1,x_2,x_3), 
\end{equation}
where operator $\Lop_\varphi^*$ is adjoint to $\Lop_\varphi$, called laminographic backprojection and written as
\begin{equation}\label{eq:adj}
    \Lop_{\varphi}^* d(x_1,x_2,x_3)= \int_0^{2\pi}\int_{\mathbb{R}^2} d(\theta,u,v)\tilde{\delta}(\varphi,\delta,u,v)\text{d}u\text{d}v\text{d}\theta .
\end{equation}
The operator $\mathcal{W}$ is described as a convolution with a transfer function being a suitable scaled version of $|\sigma|\cos(\varphi)$ (ramp filter), where $\sigma$ denotes the conjugate variable of $u$. Similar to regular tomography, instead of the ramp filter it is common to consider low pass filters (shepp-logan, parzen) for decreasing noise in reconstructions. 

Direct discretization of line integrals for evaluating the forward and adjoint laminography operators \eqref{eq:fwd} and \eqref{eq:adj} with linear interpolation has computational complexity $\mathcal{O}(N^3N_\theta)$ if we assume that reconstruction is made on a $N\times N\times N$ volume, the detector size is $N\times N$, and the number of rotation angles is $N_\theta$.

Alternatively, formulas \eqref{eq:fwd} and \eqref{eq:adj} can be evaluated with Fourier-based methods of lower computation complexity~\cite{myagotin2013efficient,voropaev2016direct}. In this work, we define the Fourier transform as
$    \mathcal{F} f (\xi) = \int f(x)e^{-2\pi i x\xi}dx$
and use subscripts with operation $\mathcal{F}$ to specify the transform dimensions or the grids it acts to. Using the Fourier transform properties of the delta function, it can be readily verified that 

\begin{equation}\label{eq:fftfwd}
    \Lop_{\varphi} \mu(\theta,u,v) = \mathcal{F}_{\text{2D}}^*\mathcal{F}_{\text{u3D}}^{\textcolor{white}{*}}\mu(\theta,u,v),
\end{equation}
where $\mathcal{F}_{\text{2D}}$ is a regular two-dimensional Fourier transform that in discrete case is computed between equally-spaced grids ($\mathcal{F}^*_{2D}$ - the adjoint/inverse transform). Operator $\mathcal{F}_{\text{u3D}}^{\textcolor{white}{*}}$ denotes the three-dimensional Fourier transform that in discrete case is applied from the equally-spaced grid $(x_1,x_2,x_3)$ to unequally-spaced grid $(\xi_1,\xi_2,\xi_3)$ with 
\begin{equation}\label{eq:uneqspace}
    \begin{aligned}
    & \xi_1 = k_u\cos_\theta+k_v\sin\theta\sin\varphi, \\
    & \xi_2= k_u\sin_\theta-k_v\cos\theta\sin\varphi, \\
    & \xi_3 = k_v\cos\varphi.
    \end{aligned}
\end{equation}
By making use of properties of the Fourier transform, the adjoint laminography operator can be calculated with replacing the 2D and 3D Fourier transforms by their adjoints and with reversing the operators order, namely,

\begin{equation}\label{eq:fftadj}
    \Lop^{*}_{\varphi} d(x_1,x_2,x_3) = \mathcal{F}^{*}_{\text{u3D}}
    \mathcal{F}^{\textcolor{white}{*}}_{\text{2D}}d(x_1,x_2,x_3),
\end{equation}

Computational complexity for discrete evaluating the forward and adjoint laminography operators by \eqref{eq:fftfwd} and \eqref{eq:fftadj} in terms of FFT is lower than by using direct discretization of line integrals in \eqref{eq:fwd} and \eqref{eq:adj}. Indeed, the two-dimensional Fourier transform between equally-spaced grids in both formulas is directly computed by means of FFT. For the three-dimensional Fourier transform $\mathcal{F}_{\text{u3D}}$ there also exist fast methods based on unequally-spaced Fast Fourier Transform (USFFT)~\cite{dutt1993fast, beylkin1998applications}. In short, the methods utilize a Gaussian function $\psi$ exhibiting certain properties, to rewrite the transform in the form of convolution,
\begin{equation}
    \mathcal{F}_{\text{u3D}}\mu(\xi_1,\xi_2,\xi_3) =  \mathcal{F}_{\text{3D}}\left(\frac{\mu}{\psi}\psi\right)(\xi_1,\xi_2,\xi_3) = \mathcal{F}_{\text{3D}}\tilde{\mu}*\mathcal{F}_{\text{3D}}\psi(\xi_1,\xi_2,\xi_3),
\end{equation}
where $\tilde{\mu} = \frac{\mu}{\psi}$. For the discrete version, the Fourier transform $\mathcal{F}\tilde{\mu}$ is calculated on an equally-spaced grid, and the convolution allows switching to unequally-spaced coordinates. The whole 3D USFFT procedure for computing the Fourier transform from equally to unequally spaced grid is described by the following steps:
\begin{enumerate}
\item{Division by $\psi$ in the space domain}
\item{3D FFT}
\item{Convolution-type operation in the frequency domain}
\end{enumerate}
The adjoint laminography operator is computed with the inverse 3D USFFT from the unequally to equally spaced grid. For that, the steps above should be done in reverse order, and the second step is replaced to 'inverse 3D FFT'. The resulting computational complexity for evaluation of the forward/adjoint laminography operator is given by the complexity of 3D FFT, i.e. $\mathcal{O}(N^3\log N)$.
Clearly, the Fourier-based method is computationally more favorable than the direct discretization of the line integral. However, for small data sizes and if the number of projection angles ($N_\theta$) is very small then the direct discretization may work faster due to the code implementation.

\begin{figure}
    \centering
    \includegraphics[width=1\textwidth]{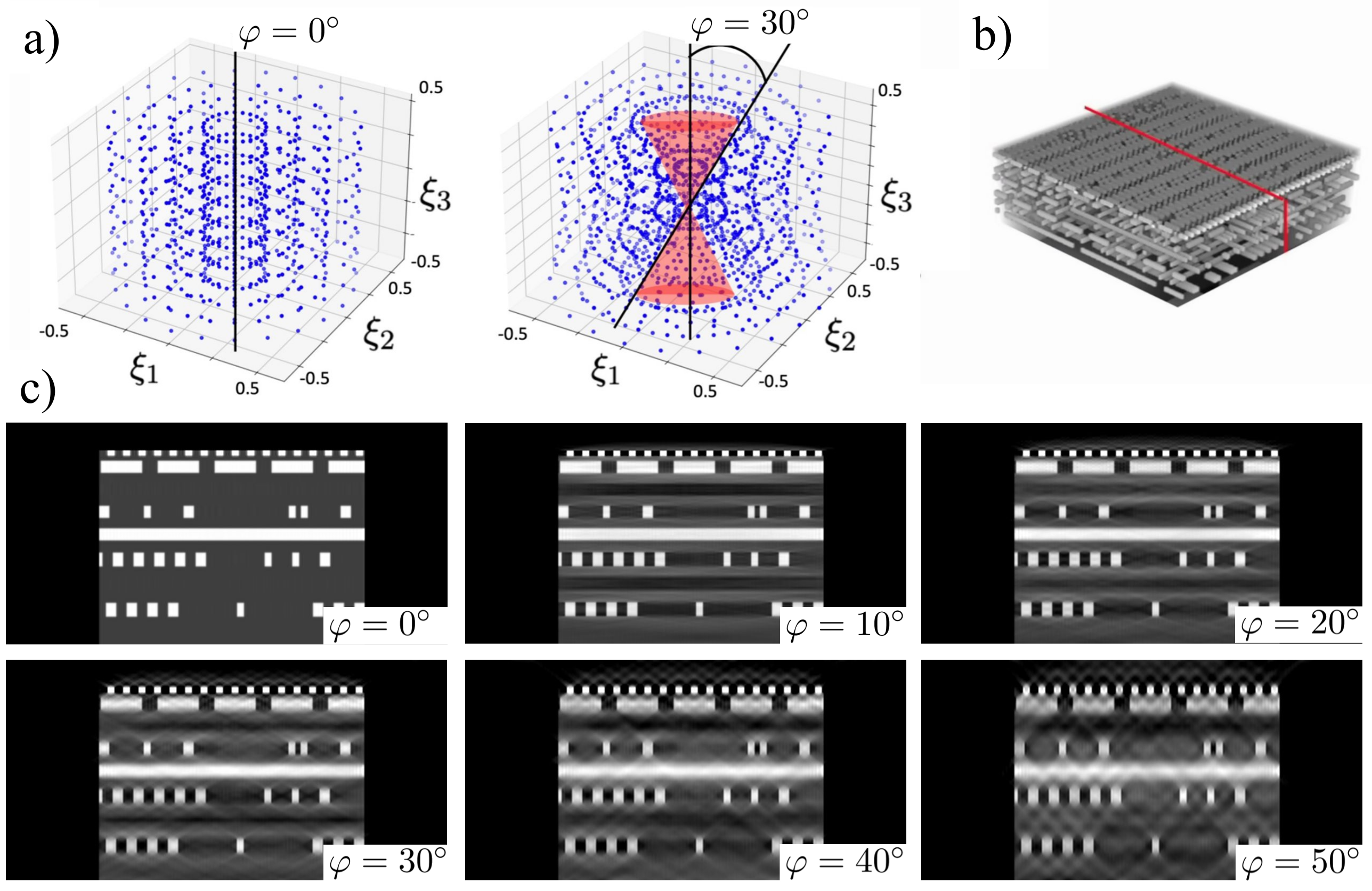}
    \caption{Undersampling problem in laminography: a) grids $(\xi_1,\xi_2,\xi_3)$ in the Fourier space defined in \eqref{eq:uneqspace} for $\varphi=0^\circ$ (regular tomography) and for $\varphi=30^\circ$ showing the missing cone in red; b) synthetic integrated circuit dataset with the red-colored position of the vertical slice for demonstrating reconstruction quality, c) examples of vertical slice reconstruction in laminography for different tilt angles.}
    \label{fig:undersamplng}
\end{figure}
Formulation of the reconstruction problem with the Fourier-based method can be used to demonstrate the general laminography undersampling problem, i.e. the missing cone in the Fourier space, see Figure~\ref{fig:undersamplng}a. The figure depicts grids $(\xi_1,\xi_2,\xi_3)$ in the Fourier space for laminography angles $\varphi=0^\circ$ (regular tomography) and for $\varphi=30^\circ$ where the region marked with red cone corresponds to missing information. To demonstrate the effect of the missing cone on reconstruction we generated laminography data for a synthetic integrated circuit data set by using formula~\eqref{eq:fftfwd} and reconstructed it by using formulas~\eqref{eq:recon} and~\eqref{eq:fftadj}. Figure~\ref{fig:undersamplng}b shows a 3D volume rendering of the integrated circuit, and an approximate red-colored position of the reconstructed slice used for quality comparisons. Figure~\ref{fig:undersamplng}c shows the reconstructed slice for the tilt angles $0^\circ,10^\circ,20^\circ,30^\circ,40^\circ$ and $50^\circ$. The "tail" artefacts along the angular span are clearly visible, especially for $40^\circ$ and $50^\circ$ tilt angles. We also observe that most of visible artefacts originate from the sample features having higher amplitudes (metal layers of white color) and propagate to the regions of uniform intensity. 
Therefore, it can be inferred that better quality laminography results are obtained with lower tilt angles ($\le30^\circ$) and for more homogeneous samples.

\begin{figure}
    \centering
    \includegraphics[width=1\textwidth]{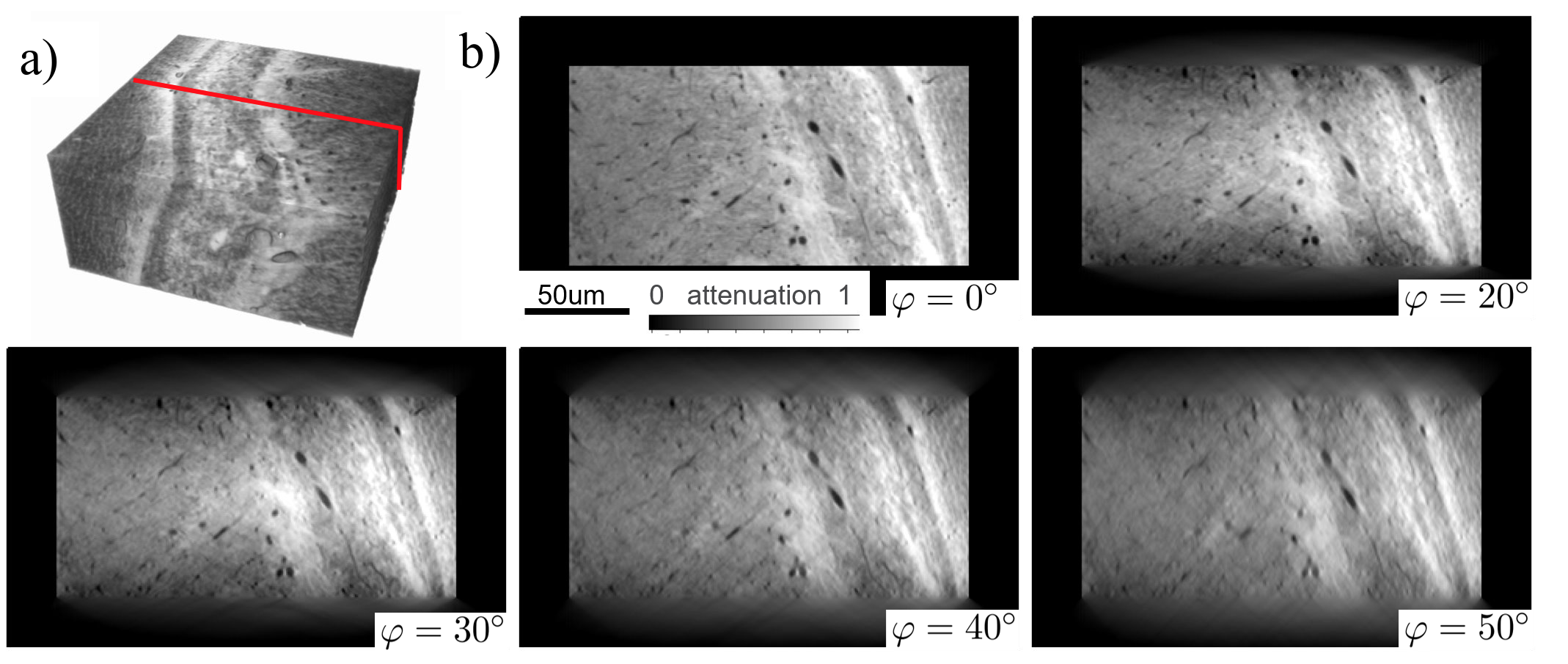}
    \caption{Reconstruction of a semi-synthetic brain dataset: a) a cropped region of the brain dataset used for data modeling; the red color shows the position of the vertical slice for demonstrating reconstruction quality, b) examples of vertical slice reconstruction in laminography for different tilt angles.}
    \label{fig:syn_brain}
\end{figure}

 In Figure~\ref{fig:syn_brain} we validate the influence of these artefacts in realistic case of scanning a section of Osmium-stained mouse brain. To demonstrate reconstruction of such sample with different laminography angles, we formed a 'semi-synthetic' mouse brain dataset based on a high-quality data purposely acquired in the computed tomography geometry. The reconstruction was then cropped to a small slab shape and used as an initial object for generating laminography data for different angles, where forward laminography operator~\eqref{eq:fwd} was used to generate projections. The figure shows that mouse brain features are not significantly affected by the artefacts for angles $\varphi\leq30^\circ$, except probably a small amplitude loose at the top and bottom parts of reconstructions. In the results with $\varphi=40^\circ,50^\circ$ the degradation of fine brain features become visible.

 The integrated circuit and brain volumes  had sizes (256,~256,~128), projection data were generated for 384 angles over 360$^\circ$ range for a simulated detector with sizes (256,~256). It is worth to note that in contrast to regular tomography operating with data from a half-circle angular range ($180^\circ$), the laminography geometry requires angles from the whole circle to properly fill the frequency space for reconstruction. 
 
 For suppressing laminography artefacts typical for the samples like the synthetic integrated circuit above, one can use reconstruction with Total Variation regularization, as it was demonstrated in~\cite{fisher2019laminography}. In Appendix A, we formulate the reconstruction problem with regularization, solve it with employing the proposed implementation of the laminography operators, and demonstrate enhancement of integrated circuit reconstruction results for large laminography angles.

\section{GPU acceleration of reconstruction}
Nvidia GPUs are commonly used for tomography reconstruction since they demonstrate more than 10$\times$ acceleration compared to CPU-based implementations~\cite{andersson2016fast,nikitin2023tomocupy}. In tomography, each data slice (sinogram) can be processed independently to obtain a horizontal slice through a 3D reconstructed object. Modern GPUs have enough VRAM memory to process sinograms of the sizes more than $30k\times30k$ pixels~\cite{nikitin2023tomocupy}.  
In laminography though, reconstruction of one slice through the object requires data from different sinograms, making the GPU memory requirements more demanding. Therefore, efficient data chunking, as well as CPU-GPU data transfer protocols, need to be developed.

For direct discretization of the line integral in the backprojection formula \eqref{eq:adj}, the chunking can be done in the slice $x_3$ and angle $\theta$ directions. Reconstructed chunk of slices in $x_3$ is obtained by summing reconstructions from all individual chunks of angles. The GPU memory requirements are defined based on discretizing  $\Lop_{\varphi}^* d(x_1,x_2,\cdot)=\int_{\mathbb{R}^2} d(\cdot,u,v)\tilde{\delta}(\varphi,\cdot,u,v)\text{d}u\,\text{d}v$, which does not involve operations on 3D arrays and therefore can be computed by chunks that fit the GPU memory. 

Reconstruction with the Fourier-based method for evaluating the backprojection by Formula \eqref{eq:fftadj} involves operations on 3D arrays. For instance, computing 3D Fourier transforms on unequally spaced grids involves computing 3D FFTs and 3D interpolation-like procedures in the frequency domain. Implementing such procedures on a GPU even for a $2048^3$ dataset requires more than 64GB memory, which is beyond the capability of most modern GPUs. Therefore, we decompose the unequally spaced 3D Fourier transform into a combination of batched 1D and 2D transforms by splitting variables in the Fourier integral as follows,
\begin{equation}\label{eq:fftfwdsplit}
\begin{aligned}
    &\mathcal{F}_\text{u3d} f (\xi_1,\xi_2,\xi_3) = \iiint e^{-2\pi i (x_1\xi_1+x_2\xi_2+x_3\xi_3) } dx_1dx_2dx_3 \\&= \iint e^{-2\pi i (x_1\xi_1+x_2\xi_2)}dx_1dx_2 \left(\int f(x_1,x_2,x_3)e^{-2\pi i x_3\xi_3}dx_3\right)
\end{aligned}
\end{equation}
It turns out that computing $\mathcal{F}_\text{u3d}f$ can be done in two steps: 1D USFFT with respect to variable $x_3$, followed by 2D USFFT with respect to variables $x_1,x_2$. Data chunking to fit GPU memory is done by columns and slices, respectively.

\begin{figure}
    \centering
    \includegraphics[width=0.8\textwidth]{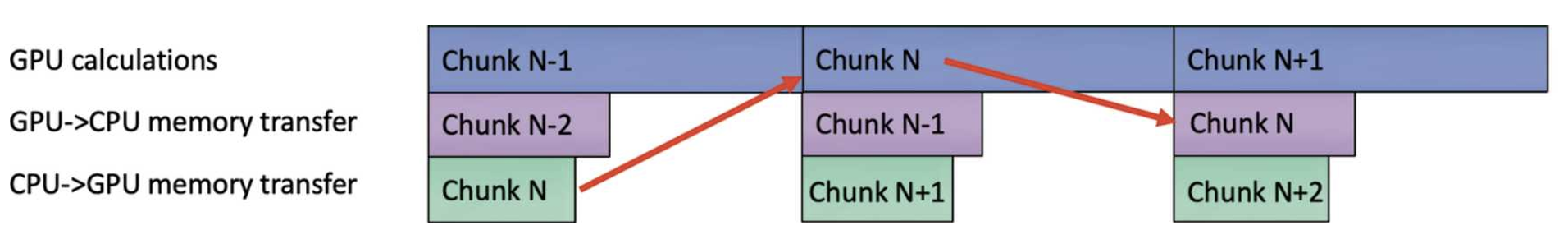}
    \caption{A scheme for asynchronous data processing by chunks where GPU reconstructions are overlapped with data transfers.}
    \label{fig:pipeline}
\end{figure}

GPU data processing by chunks involves 3 operations: CPU to GPU data transfer, computations on GPU, and GPU to CPU data transfer. In the aforementioned approach with  batched processing the CPU-GPU data transfer takes significant amount of total computation time for reconstruction. Therefore, in this work we adopt the approach proposed in ~\cite{nikitin2023tomocupy} and organize an efficient pipeline for asynchronous data processing by chunks. Schematically the pipeline is shown in Figure~\ref{fig:pipeline}. With this pipeline, three operations are executed simultaneously: CPU–GPU memory transfer for Chunk N; GPU computations for Chunk 
N - 1; and GPU-CPU memory transfer for Chunk N - 2. We used this approach in all steps that involve chunking for computing backprojection~\eqref{eq:adj}, i.e. 2D USFFT, 2D FFT, and 1D USFFT. 

Asynchronous execution of computations and fast data transfers is implemented in Python with the CuPy interface. The Cupy interface allows for creating Nvidia CUDA Streams and allocating pinned memory needed to overlap computations and data transfers. To implement the overlap, the pinned memory on CPU and device memory on GPU should be both allocated for two input data chunks and two output chunks. Three CUDA streams run simultaneously by switching between chunks: the first stream performs a data copy to the first input chunk of the pinned memory, followed by transfer to the first input chunk of GPU memory. The second stream performs GPU computations on the second input chunk in GPU memory (whenever it is available) and places the result in the second output chunk in GPU memory. The third stream executes a data transfer from the first output chunk in GPU memory to the first output pinned memory chunk. The chunk is then copied to a corresponding place in the resulting array. After processing each chunk, all streams synchronize and switch the chunk ID (0 or 1) they operate with.

\begin{figure}
    \centering
    \includegraphics[width=1\textwidth]{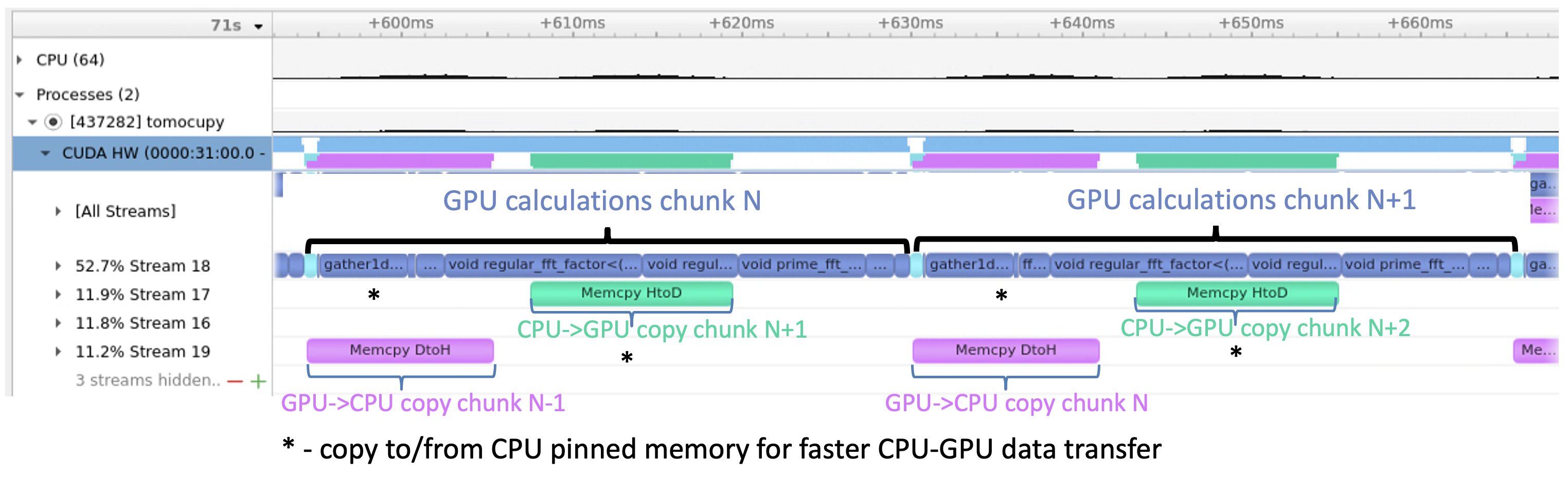}
    \caption{Timeline view report from the Nvidia Nsight System tool for asynchronous execution of the 1D USFFT operation for computing the laminographic backprojection operator.}
    \label{fig:nsight}
\end{figure}

Asynchronous execution can be verified with the Nvidia Nsight System profiling tool\footnote{https://developer.Nvidia.com/nsight-systems}. As an example, in Figure~\ref{fig:nsight} we demonstrate profiling results after executing a batch of 1D USFFT. Memory transfers take more than ~70\% of time for GPU calculations, however, they are done asynchronously and therefore do not affect performance. Note that in comparison to the schematic in Figure~\ref{fig:pipeline}, each CPU to GPU data transfer (green block) starts later than GPU data processing (dark blue block). This is due to the fact that some CPU time is spent for transferring data to pinned memory, which is referred to CPU execution and is not shown in the profiler. Similarly, some CPU time for operation with pinned memory is spent after transferring data from GPU to CPU. In Figure~\ref{fig:nsight} we marked these blocks with "*" for clarity.

In Table~\ref{tab:perf} we demonstrate performance tests of the laminographic reconstruction implemented in TomocuPy package. Besides reconstruction with the Fourier-based method (\textit{Fourierrec}) and by direct discretization of the backprojection line integral~\eqref{eq:adj} (\textit{Linerec}), the table also shows time for data pre-processing and read/write operations with an SSD storage. The pre-processing step include dark-flat field correction, ring removal~\cite{vo2018superior} and phase-retrieval with the Paganin filtering~\cite{paganin2002simultaneous}. All operations are implemented in TomocuPy using a similar GPU pipeline approach as for the backprojection.

The tests were performed using synthetic HDF5 format datasets of different sizes. The datasets were generated for $N$ 8-bit laminographic projections with $N\times N$ detector sizes, where $N$ ranges from 1024 to 4096. The laminography tilt angle was set to $20^\circ$, although this angle does not affect the performance significantly. Reconstructed volumes $(N\times N\times N)$ were obtained as sets of tiff files in 32-bit precision. Chunk sizes in projection angles and reconstructed slices were chosen by taking into account the GPU memory limitation and overall performance. For instance, for $N=1024$ the chunk size in angles was 128, while for $N=4096$ the size was 4. The chunk sizes as well as other reconstruction are given through the TomocuPy command-line interface, see the next section for more details.

    Performance tests were carried out on a machine with Intel Xeon Gold 6326 CPU @ 2.90GHz, 2TB DDR4 3200 memory, one Nvidia Tesla A100 with 40GB memory, and Intel SSD D7-P5510 Series PCIe 4.0 NVMe disks. Installed software included Python 3.9, CuPy 12.1, Nvidia CUDA toolkit 12.1.

\begin{table}[]
    \setlength\tabcolsep{4pt}
    \centering
    \begin{tabular}{|c|c|c|c|c|c|c|c|c|c|}
    \hline
         N& Data read & Pre-processing & \multicolumn{3}{c|}{Reconstruction}& Recon write & \multicolumn{2}{c|}{Total}   \\
         &(8-bit)&& Fourierrec& Linerec & Gain & (32-bit)& Fourierrec& Linerec\\
         \hline
         1024& 1.3 & 1.9 & 3.0& 9.3 & 3.1 & 4.2 & 10.4 & 16.7\\
         1536& 2.3 & 4.7 & 8.6& 52.1 & 6.1 & 8.0& 23.6 & 67.1\\
         2048& 4.6 & 7.8 & 21.1& 222.1 & 10.5 & 18.1& 51.6 & 252.6\\
         3072& 13.5 & 27.4 & 77.7& 1002.2 & 12.9 & 52.3 & 170.9 & 1095.4\\         
         4096& 31.1 & 59.4 & 164.0& 4937.5 & 30.1 & 120.3 & 374.8 & 5148.3\\         
         \hline
    \end{tabular}
    \caption{Time in seconds for processing $N$ laminographic projections of size $N\times N$ to reconstruct $N\times N\times N$ volumes: "Data read" - parallel read of 8-bit data from an SSD storage, "Pre-processing" - dark-flat field correction, ring removal, Paganin filtering, "Reconstruction" - reconstruction with the the Fourierrec (Fourier-based) method, and with Linerec (direct discretization of the line integral) method, "Recon write" - parallel write of 32-bit reconstructions to the SSD storage.}
    \label{tab:perf}
\end{table}

Table~\ref{tab:perf} shows the Fourierrec method significantly outperforms Linerec because of more favorable computational complexity ($\mathcal{O}(N^3\log N)$ vs $\mathcal{O}(N^4)$). For $N=1024$ the acceleration is about $3\times$, while for $N=4096$ it is higher than $30\times$. Lower computational complexity is crucial in developing new reconstruction algorithms because detector sizes become bigger. Even if an algorithm with $\mathcal{O}(N^4)$ complexity is accelerated with a large computational resources like multiple GPUs, for a large enough $N$ it will become slower than the one with $\mathcal{O}(N^3\log N)$ complexity. From the table, we see that increasing data sizes by a factor of 2 (e.g. 2048$\to$4096) gives difference between reconstruction times as $2^4=16$ for the Linerec method, and  $2^3=8$ for the Fourierrec method. However, we also observe that increasing data sizes by a factor of 1.5 (e.g. 3072$\to$4096) does not give $1.5^3\approx3.4$ time difference (77.7~s vs 164~s). This can be explained by the fact that the FFT procedure on GPU is very well optimized for the sizes that are powers of 2, making processing data for $N=4096$ more optimal. Simple test of the two-dimensional FFT operation on GPU gives execution times 0.17~ms, 0.43~ms, 0.5~ms for sizes $2048\times2048$, $3072\times3072$, $4096\times4096$, respectively, which confirms the slowdown for the sizes that are not powers of 2.     

Another observation from Table~\ref{tab:perf} is relatively high performance of read and write operations. In these tests, we utilized PCIe 4.0 NVMe SSDs that allow parallel operations with the storage using multi-threading. With this system we were able to reach up to 3~GB/s for reading and writing. Regular HDD storage is more than 5 times slower and therefore may become a bottleneck for reconstruction. Therefore we note that, besides powerful GPUs, the NVMe SSD storage is also a crucial component for accelerating the whole reconstruction process. The table demonstrates results for up to $4096\times4096\times4096$ reconstructed volumes, which corresponds to 256~GB of RAM. For bigger sizes the data may not fit into RAM and it will be necessary to operate with data chunks by communicating with the hard disk for each pre-processing procedure and reconstruction step (e.g., Paganin Filter, USFFT1D, USFFT2D, etc.). In this case a fast SSD storage could be a good alternative to RAM. 

\section{Application to neuroimaging}

\begin{figure}
    \centering
    \includegraphics[width=1\textwidth]{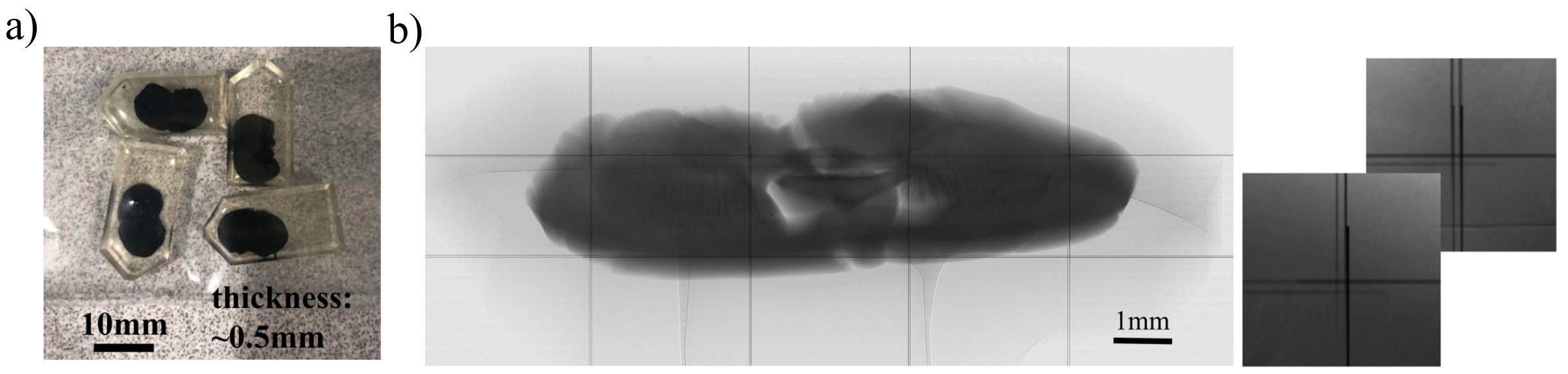}
    \caption{Mouse brain samples used for laminographic scanning (a) and an example of laminographic projection after data stitching and dark-flat field correction (b).}
    \label{fig:brainslabs}
\end{figure}

To demonstrate applicability of the developed laminographic implementation for imaging large samples we considered imaging of 4 sequential sample slabs cut from a whole mouse brain, see Figure~\ref{fig:brainslabs}a. 

The slabs were prepared using protocols for electron microscopy. Briefly, a mouse is transcardially perfused with aldehyde fixatives and sectioned into $\sim$500 $\mu m$ coronal sections. Sections are then stained with heavy metals (i.e., osmium tetroxide, uranyl acetate, and lead (II) nitrate)~\cite{hua2015large} to increase the X-ray absorption contrast. After staining, the samples were dehydrated and embedded in an epoxy  resin to make it more X-ray resistant. Each slab has approximate sizes 12~mm~$\times$~8~mm, with the thickness of about 500~$\mu m$. For 25~keV energy (optimal for 2-BM beamline of APS) such thickness indicates satisfactory X-ray transmission, ranging from 15\% to 30\% depending on the sample rotation angle. The sample slabs were glued to a Kapton\textsuperscript{\textregistered} tube which in turn was attached to a kinematic mount with the Bondic\textsuperscript{\textregistered} UV glue, see Figure~\ref{fig:samplemount}. Experiments were conducted in the downstream experimental station located at 50~m from the source. For the measurements we used a filtered pink X-ray beam with the energy peak at 25~keV. A 8~mm glass filter was placed upstream of the sample to cut low X-ray energies and decrease the radiation damage. The exposure time per projection was 50 ms. ORX-10G-310S9M camera with $6464\times4852$ pixels (pixel size 3.45~$\mu$m~$\times$~3.45~$\mu$m) recorded projections from a 25~$\mu$m-thick GGG:Eu scintillator, magnified through a 7.5$\times$ lens yielding a resulting isometric voxel size of 0.92~$\mu$m after $2\times2$ binning. We experimentally chose 200~mm as the distance between the sample and the objective to increase propagation-based phase contrast.  The detector field of view after binning was also cropped to the size $3232\times2256$ due to the beam shape. Detector data were collected in 12-bit mode and stored as 16-bit images.  Laminography tilt angle was set to 20$^\circ$ by adding the 10$^\circ$ wedge under the rotary stage and setting the tilt motor to 10$^\circ$, see Figure~\ref{fig:setup2bm}.

The mouse brain slabs were scanned in the mosaic scanning mode by moving the whole sample stack by 5 and 3 steps in horizontal (orthogonal to the beam) and vertical and directions, respectively. Because of low repeatability and accuracy of the vertical and horizontal motors under the rotary stage, overlap of 300-400 pixels between projections of two adjacent datasets was set to perform image registration for more precise image stitching. For feature-based image registration we used the SIFT algorithm~\cite{lowe2004distinctive}. Data from each two overlapped regions were summed with linearly changing weights in the range between 0 and 1. An example of one the stitched projections after dark and flat field correction is shown in Figure~\ref{fig:brainslabs}b. Black lines indicate borders between different datasets use for stitching. Projection size after stitching is 14960$\times$5936 pixels.

15000 tomographic projections were collected in a fly scan mode while the sample was continuously rotated over 360$^\circ$ at 0.48~$^\circ/s$, which together with collection of dark and flat fields yielded 13 min total acquisition time per dataset, 5$\times$3$\times$13~=~195~min per a slab, and 4$\times$5$\times$3$\times$13 = 780~min (13~h) for scanning all 4 slabs. Total size of acquired raw data was about 12~TB, after projections stitching it reduced to 10~TB. 

For reconstruction we used the FBP formula \eqref{eq:recon} with the backprojection operator implemented with  the proposed Fourier-based method, see Formula~\eqref{eq:fftadj}. The iterative reconstruction was not considered since laminography artefacts for 20$^\circ$ tilt angle are not significant for such kind of samples (see tests in Figure~\ref{fig:syn_brain}). Because of huge data sizes, reconstruction was done by steps with saving and loading intermediate results for chunked data processing.

To accelerate reconstruction, we utilized several nodes of the Polaris supercomputer of the Argonne Leadership Computing Facility (https://www.alcf.anl.gov/polaris). Each Polaris node is equipped with an AMD EPYC Milan processor and four Tesla A100 GPUs with the SXM connection interface and high-speed HBM memory architecture. The storage called Eagle is based on a Lustre file system residing on an HPE ClusterStor E1000 platform equipped with 100 Petabytes of usable capacity across 8480 disk drives. This ClusterStor platform also provides 160 Object Storage Targets and 40 Metadata Targets with an aggregate data transfer rate of 650GB/s. 

\subsection{Reconstruction pipeline with TomocuPy calls}
Before demonstrating reconstruction of large mouse brain slabs, we will describe our proposed laminographic pipeline for manual adjustments of the rotation axis and laminographic tilt angle by using a dataset acquired for a small mouse brain slab that almost fits the detector field of view. The dataset consists of 3000 projections of the size $3232\times2256$. 

In the proposed laminography implementation, the tilt angle is not given exactly since the wedge is manufactured with low angle accuracy. Moreover,   
the wedge may have a error in the roll angle, i.e. the one in the direction orthogonal to the beam. The roll angle issue can be resolved with the regular tomographic setup: by moving the tilt angle to 0, rotating camera or the roll motor under the rotary stage, moving the tilt angle back to 20$^\circ$. 

\begin{figure}
    \centering
    \includegraphics[width=1\textwidth]{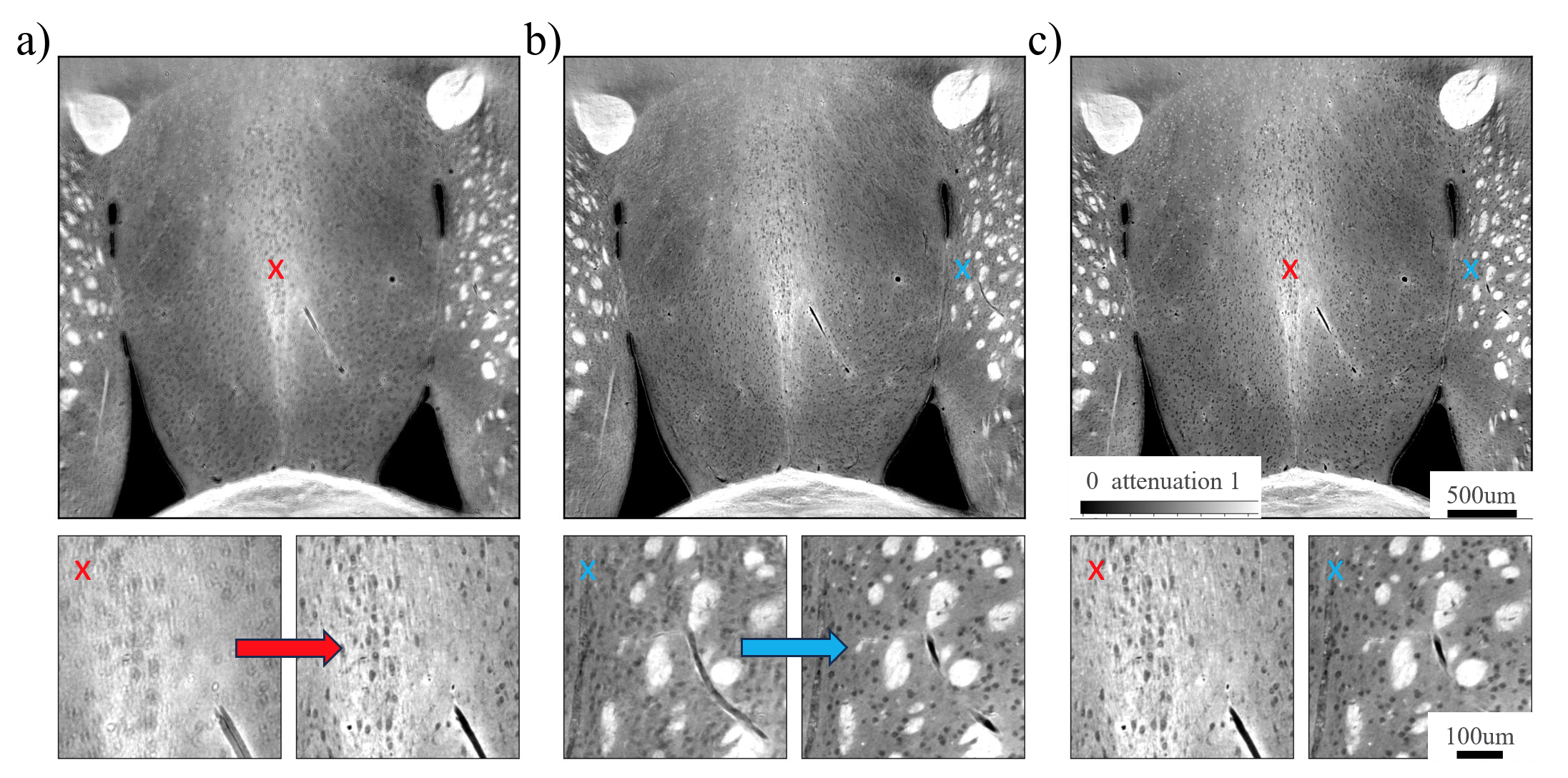}
    \caption{Reconstruction strategy for the low-cost laminography: a) Step 1, searching the rotation axis by examining the middle part of the image, b) Step 2, searching the laminography tilt angle by setting the rotation axis from Step 1 and examining the border part of the image, c) Step 3, full reconstruction with chosen rotation axis and laminography tilt on Steps 1 and 2.}
    \label{fig:recbrain}
\end{figure}

Searching for the laminography tilt angle and searching for the rotation axis can be performed during reconstruction.
We propose the following strategy:

\begin{enumerate}
    \item[Step 1.] Choose an approximate value for the laminography tilt angle and for the rotation axis, and run reconstruction of 1 slice for different rotation axes. In TomocuPy command-line interface the command should include parameter \verb|--reconstruction-type try| and is executed as:
    \begin{verbatim}    
tomocupy recon_steps --file-name --file-name brain_x2y1.h5  
--reconstruction-type try --rotation-axis 1616 --center-search-width 20 
--center-search-step 0.5 --lamino-angle 20 --nsino 0.3
    \end{verbatim}  
    \vspace{-1cm}
    The command generates reconstructions of 1 slice  for rotation axes $[1616-20,\dots,1616+20)$ and for 20$^\circ$ laminography tilt angle. Reconstruction of the middle part of images is not influenced much by the error in the laminography tilt angle, therefore the rotation axis can be found by scrolling through the images and examining only the middle part of them, see Figure~\ref{fig:recbrain}a.
    \item[Step 2.] Choose an approximate value for the laminography tilt angle, set the rotation axis found in Step 1 and run reconstruction of 1 slice for different laminography tilt angles. In Tomocupy command-line interface the command should include parameter \verb|--reconstruction-type try-lamino| and may be executed as:
    \begin{verbatim}    
tomocupy recon_steps --file-name brain_x2y1.h5 
--reconstruction-type try-lamino --rotation-axis 1630.5 --lamino-search-width 2
--lamino-search-step 0.1 --lamino-angle 20 --nsino 0.3
    \end{verbatim}  
    \vspace{-1cm}
    The command generates reconstructions of 1 slice for rotation axis 1630.5 and for $[20-2,\dots,20+2)^\circ$ laminography tilt angles. Reconstruction of the border parts of the images is influenced by the error in the laminography tilt angle, therefore the angle can be found by scrolling through the images and examining their border parts, see Figure~\ref{fig:recbrain}b.
    \item[Step 3.] Use the rotation axis found in Step 1 and the laminography tilt angle from Step 2 to run reconstruction of the full volume with setting parameter \verb|--reconstruction-type full|:
    \begin{verbatim}    
tomocupy recon_steps --file-name brain_x2y1.h5 
--reconstruction-type full --rotation-axis 1630.5 --lamino-angle 19.95 
    \end{verbatim}  
The command generates reconstruction of the full volume, see Figure~\ref{fig:recbrain}c. 
\end{enumerate}
Additional phase-retrieval procedure with the Paganin filter is performed by adding parameters \verb|--retrieve-phase-method, --propagation-distance, ...|, for details see TomocuPy documentation. 

\begin{figure}
    \centering
    \includegraphics[width=0.6\textwidth]{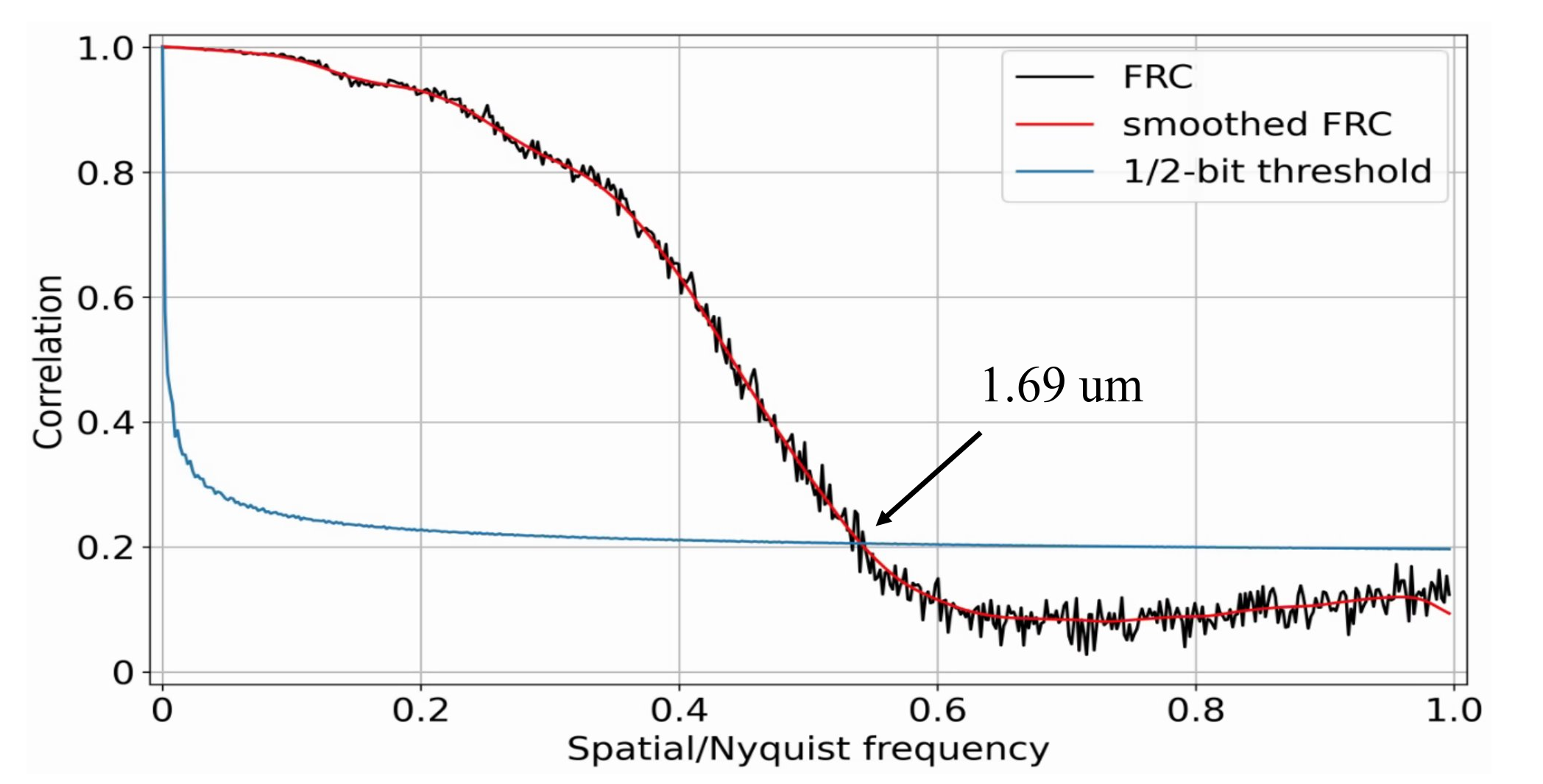}
    \caption{Resolution estimation by the Fourier Ring Correlation with the 1/2-bit criterion.}
    \label{fig:frc}
\end{figure}

Resolution levels were estimated by computing the Fourier
ring correlation (FRC)~\cite{van2005fourier} between reconstructions obtained from two independent sets of 3000 projections. We used the the 1/2-bit resolution criterion. Since the reconstructed volume is thin, the resolution levels were estimated by slices and the plot with the lowest resolution (higher level in microns) was chosen as the final result, see Figure~\ref{fig:frc}. The intersection between the lines for the 1/2-bit criterion and the FRC corresponds to 1.69~$\mu m$ resolution estimation.

\subsection{Results for the large mouse brain data}

\begin{figure}
    \centering
    \includegraphics[width=1\textwidth]{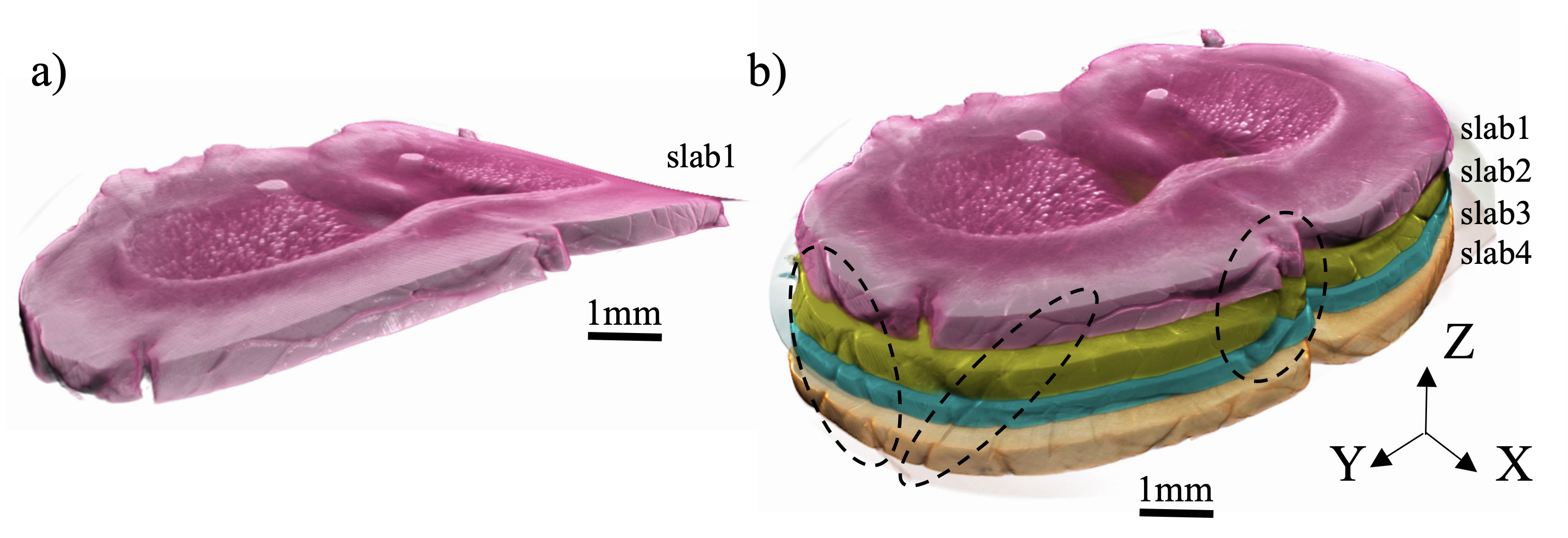}
    \caption{3D visualization of reconstructed mouse brain slabs: a) initial bent reconstruction of the slab 1 (the top sample part), b) 4 straighten slab volumes stitched together.}
    \label{fig:brain3d}
\end{figure}

In what follows we will demonstrate reconstruction of full brain slabs where data for each slab has sizes 15000$\times$14960$\times$5936 in 16 bit precision. Figure~\ref{fig:brain3d} shows 3D volume rendering of reconstructed slabs in ORS Dragonfly package after binning reconstruction by a factor of 8 in each dimension. The brain slabs were bent and tilted during the sample preparation, see Figure~\ref{fig:brain3d}a. Therefore we used additional postprocessing procedures to straigthen reconstructed volumes. The images were rotated and unbent with \textit{Image.rotate()} and \textit{Image.distort()} methods from \textit{Ward} python package. After straightening the slabs they were stitched together, see Figure~\ref{fig:brain3d}b. Black dashed ellipses in the figure show matching features between adjacent slabs.

\begin{figure}
    \centering
    \includegraphics[width=1\textwidth]{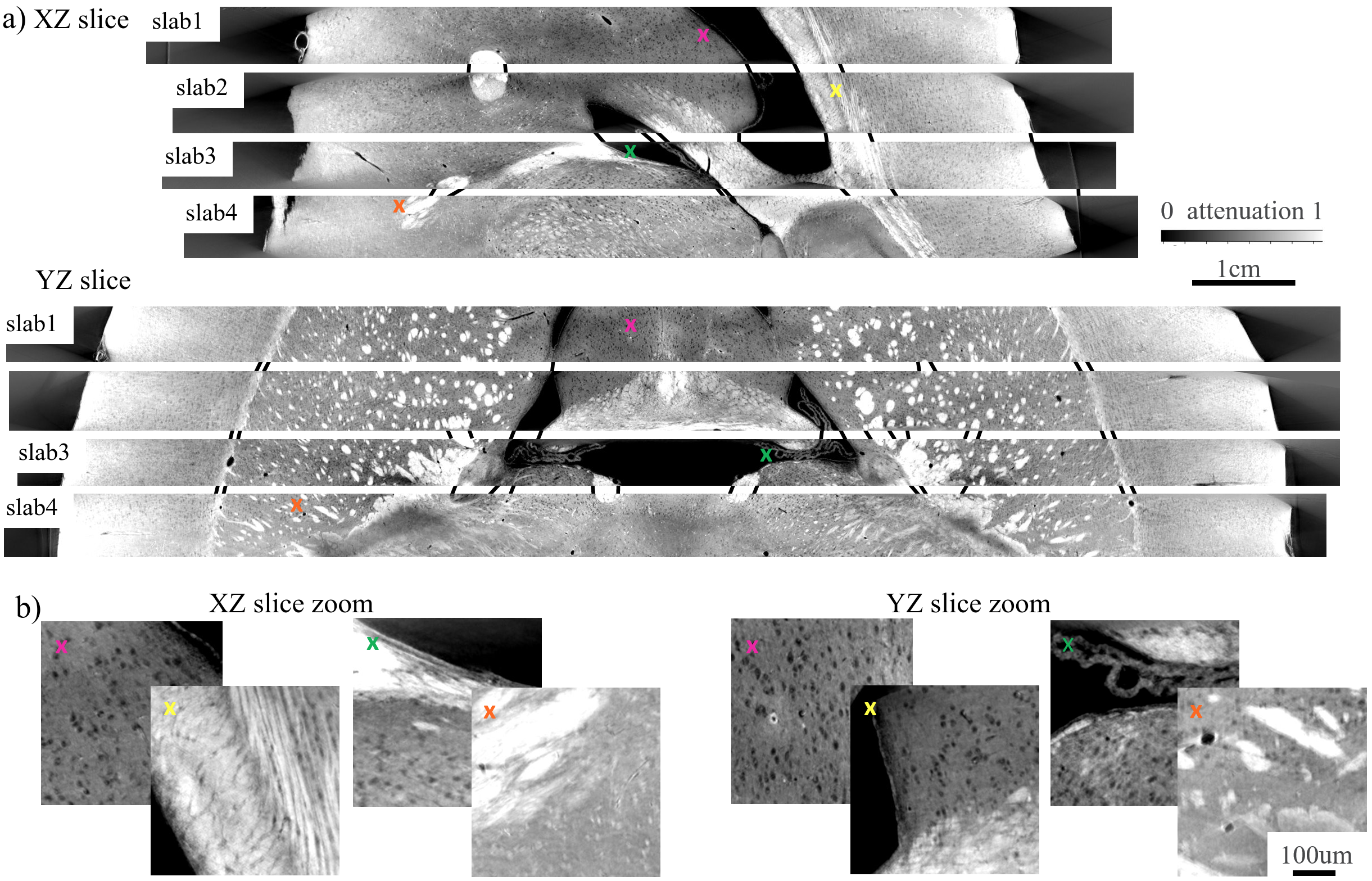}
    \caption{Stitching reconstructed mouse brain slabs in vertical directions XZ and YZ  (a), corresponding zoomed-in regions marked with colored crosses (b).   }
    \label{fig:xzyz}
\end{figure}

\begin{figure}
    \centering
    \includegraphics[width=0.9\textwidth]{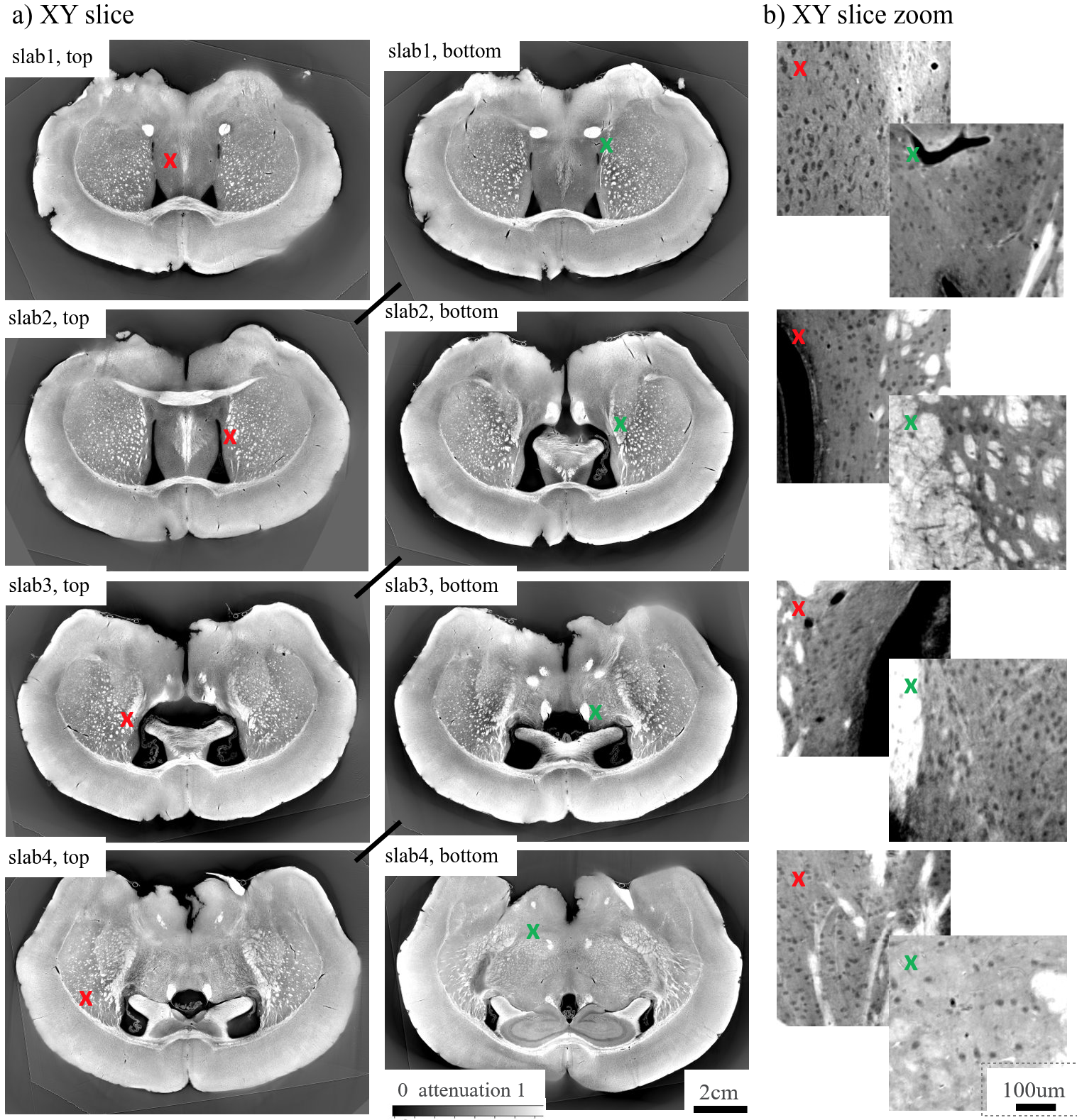}
    \caption{Top and bottom slices in horizontal direction XY for each slab (a), corresponding zoomed-in regions marked with colored crosses (b). }
    \label{fig:xy}
\end{figure}

Figures~\ref{fig:xzyz} and \ref{fig:xy} show reconstructed slices through the whole 3D volumes in high resolution. XZ, YZ, and XY directions for the slices are defined based on the axes depicted in the bottom part of Figure~\ref{fig:brain3d}. 
We carried out a visual inspection of slices for different slabs and found similar features that can be used for stitching. Black lines between the slices in Figure~\ref{fig:xzyz}a show possible connections between the features. Accurate stitching is not possible because parts of the sample were destroyed/bent due to the cutting procedure.
Based on the reconstructions, we can assume that 100$\mu m$-thick layers between slabs were destroyed while cutting. It should be also noted that the brain was cut before embedding with petropoxy. This procedure may also affect the slab structures. The missing layer due to cutting can be also observed by comparing top and bottom slices in the XY direction in Figure~\ref{fig:xy}a. 
The slices that should look similar are connected with black lines. For instance, 'slab1, bottom' and 'slab2, top' have similar features, although they are not very close to each other. Some parts of the top and bottom slices are blurred, see for instance the bottom right part of 'slab1, top', or the top right of 'slab3, bottom'. This is due to the bent structure of the slabs. Although the images were straighten after reconstruction using the \textit{Ward} python package, local deformation is not easily compensated. 
The imaging quality can be analyzed using the zoomed-in regions demonstrated in Figures~\ref{fig:xzyz}b and ~\ref{fig:xy}b. The regions are taken at the positions indicated by colored crosses in the whole slab images. The obtained imaging resolution allows for segmenting axons (black dots) in most places.   

Resolution levels were estimated with the FRC as it was shown earlier for a small mouse brain sample. The middle part of the sample confirms 1.6-1.7~$\mu m$ resolution. The levels on the sample borders are lower due to radiation damage. Iterative approaches with compensating sample deformation, e.g. the one from~\cite{nikitin2021distributed}, may be further considered to improve image quality and resolution.


\section{Conclusions and outlook}
The proposed laminographic scanning strategy, coupled with an innovative laminography instrument setup at 2BM, APS, and advanced reconstruction capabilities integrated into the TomocuPy package, not only facilitates the scanning of flat-shaped samples but also showcases the potential for imaging larger samples with minimal cutting procedures. As a primary illustration, we focused on imaging four sequential slabs from an entire mouse brain sample with a measured resolution of 1.69~$\mu m$ (0.92 micrometers voxel size). This approach allowed us to trace connections between the slabs and discern axons in high-resolution reconstructions.

While the laminography imaging technique is already established at the bending magnet beamline 2-BM of the Advanced Photon Source, our work suggests several avenues for significant improvement in brain imaging quality. First and foremost, the development of more refined cutting mechanisms is imperative. In our current sample preparation, the missing layer destroyed during cutting was approximately 100 micrometers, complicating accurate axon tracing or making it impossible in some instances. One can potentially consider methods used in electron microscopy where the destroyed layer could be less than a micron~\cite{mikula2015high}. Another possible method involves the development of methods of sectioning the brain prior to staining with heavy metals. Indeed, we have recently shown that a new machine used for sectioning aldehyde-fixed brains, called the Compresstome, has an estimated tissue loss between sections of approximately 680nm~\cite{wildenberg2023pipeline}. Given that $\sim$680 nm is near the size of a single pixel in our measurements, it is possible that the loss could be even smaller. Such an approach also offers the advantage that staining whole brains with heavy metals is difficult due to their poor diffusion, and protocols have only been demonstrated on whole mouse brains. Sectioning the tissue first and then staining it for x-ray imaging would bypass this limitation and pave a pathway towards imaging arbitrarily large brains.

Addressing the issue of the cut section deformations presents another potential enhancement. This can be approached by either performing cutting into slabs after embedding the entire sample in petropoxy or by considering more advanced methods for straightening the slabs. For example, the warp filtering method proposed in~\cite{ju20063d} for dealing with wavy histological mouse brain sections in optical microscopy can potentially be adapted for 3D X-ray images.

The challenge of projection stitching for mouse brain data arises from low contrast, leading to insufficient features for accurate alignment. To overcome this, future efforts will involve the incorporation of high-contrast patterns placed in the beam before and after scanning each slab position. These patterns are expected to enable stitching accuracy of less than 1 micron.

Additionally, we observed sample deformation at the borders due to radiation damage with the current pink X-ray beam centered at 25 keV. The monochromatic beam would be a more suitable choice. The upcoming Upgrade of the Advanced Photon Source will provide the opportunity to work with higher energies (40-50 keV) at the bending magnet beamline 2-BM. This advancement should enable the imaging of thicker slabs with reduced radiation damage.

After successfully obtaining micron-resolution laminographic images of the mouse brain, our attention now shifts towards advancing techniques for nanometer resolution. This could involve utilizing the Projection X-ray Microscope instrument planned for construction with the APS Upgrade~\cite{bean2021}. By cutting mouse brains into thinner slabs and handling significantly larger data volumes, these steps represent crucial progress towards achieving high-resolution imaging of the human brain.

\ack{Acknowledgements}

This research used resources of the Advanced Photon Source Argonne Leadership Computing Facility, U.S. Department of Energy (DOE) Office of Science User Facilities operated for the DOE Office of Science by Argonne National Laboratory under Contract No. DE-AC02-06CH11357. The author also acknowledges support from ANL's Laboratory Directed Research and Development (LDRD) funding 2021-0173, 2023-0104, 2023-0108.

\referencelist     

\appendix
\section{Reconstruction with regularization}
In this Appendix we demonstrate a scheme for laminographic reconstruction with TV regularization. The scheme allows for suppressing laminography artefacts and improving quality of results for the samples having a lot of features that are significantly different in amplitudes. We will consider the augmented Lagrangian formulation of the reconstruction problem and solve the problem by using ADMM~\cite{Boyd:11} with splitting the whole problem by local sub-problems.

As in the previous section, let $\mu(x_1,x_2,x_3)$ is a three-dimensional object and $d(\theta,u,v)$ its laminography data. Then the reconstruction problem with TV regularization reads as

\begin{equation}
\label{eq:mainregproblem}
    \frac{1}{2}\|\Lop \mu-d\|_2^2+\alpha\|\nabla \mu\|_1 \to \min,
\end{equation}
with
\begin{equation}
\alpha\|\nabla \mu\|_1=\alpha\left\lVert\sqrt{\left(\frac{\partial \mu}{\partial x}\right)^2+\left(\frac{\partial \mu}{\partial y}\right)^2+\left(\frac{\partial \mu}{\partial z}\right)^2}\right\rVert_1, \end{equation}
where parameter $\alpha$ controls the trade-off between the data fidelity and regularization terms. The TV regularization promotes sparseness in reconstructions, resulting in noise suppression and incompleteness artefacts reduction~\cite{Chambolle:16}. In particular, the 'tail' laminography artefacts is an example of data incompleteness in the frequency domain.

To solve the minimization problem~\eqref{eq:mainregproblem} we first reformulate it as an equivalent constraint optimization problem with a new auxiliary variable $\psi$,
\begin{equation}\label{Eq:minF}
\begin{aligned}
    & \min_{\mu, \psi}     & &\frac{1}{2}\|\Lop \mu-d\|_2^2+\alpha \|\psi\|_1  \\
    & \text{subject to}        & & \nabla \mu = \psi,
\end{aligned}
\end{equation}
and try to minimize the augmented Lagrangian written as follows,
\begin{equation}\label{Eq:alagr}
  \mathcal{A}_{\rho} (\mu, \psi, \lambda) =
  \frac{1}{2}\|\Lop \mu-d\|_2^2+\alpha \|\psi\|_1 + \lambda^T(\nabla \mu-\psi)+\frac{\rho}{2}\|\nabla \mu-\psi\|_2^2,
\end{equation}
where $\rho>0$ is a penalty parameter and $\lambda$ represents the dual variable. We use ADMM to split the minimization problem of the augmented Lagrangian into two local sub-problems with respect to $\mu$ and $\psi$. The sub-problems are then coordinated through variable $\lambda$ to find a solution for the original problem. Specifically, the following steps are performed in each ADMM iteration $k$:
\begin{align}
  &\mu^{k+1} =      \argmin_{\mu}      \mathcal{A}_{\rho}\left(\mu,\psi^k,\lambda^k\right)\!,\label{Eq:tomo}\\
  &\psi^{k+1} =  \argmin_{\psi} \mathcal{A}_{\rho}\left(\mu^{k+1},\psi, \lambda^{k}\right)\!,\label{Eq:reg}\\
  &\lambda^{k+1} = \lambda^k + \rho \left(\nabla \mu^{k+1} -\psi^{k+1}\right)\!,\label{Eq:lamd}
\end{align}
for zeros or some adequate initial guess at $k=0$. 

The minimization functional for the problem~\eqref{Eq:tomo} can be written as follows,
\begin{equation}\label{Eq:tomoe}
  F(\mu)= 
  \frac{1}{2}\|\Lop \mu-d\|^2_2+ \frac{\rho}{2}\|\nabla \mu-\psi^k+\lambda^{k}/\rho\|_2^2,
\end{equation}
where the terms not depending on $\mu$ are dropped. The problem is solved by considering the steepest ascent direction $\nabla_\mu F(u)$ given as,
\begin{equation}
  \nabla_\mu F(\mu) = \Lop^T\!(\Lop \mu -d)-\rho\,\text{div}(\nabla \mu -\psi^k+\lambda^k/\rho),
  \label{Eq:gradtomo}
\end{equation}
where the divergence operator $\text{div}$ is the adjoint to $-\nabla$.
With the steepest ascent direction, we can construct iterative schemes for solving \eqref{Eq:tomo} by using methods with different convergence rates. Here we employ the Conjugate Gradient (CG) method for its faster convergence rate at the expense of memory requirements. CG iterations are given as $u_{m+1} = u_m + \gamma_m \eta_m$,
where $\gamma_m$ is a step length computed by a line-search procedure~\cite{nocedal2006numerical} and $\eta_m$ is the search direction that we compute by using the Dai-Yuan formula~\cite{DaiYuan:99},
\begin{equation}\label{Eq:DaiYuan2}
\begin{aligned}
  \eta_{m+1}\!=\!-\nabla_\mu F(\mu_{m+1})\!+\!\frac{\|\nabla_\mu F(\mu_{m+1})\|_2^2}{(\nabla_\mu F(\mu_{m+1})\!-\!\nabla_\mu F(\mu_{m}))^T\eta_m}\eta_m,
  \end{aligned}
\end{equation}
where $\eta_0=-\nabla_\mu F(\mu_0)$.
On each ADMM iteration, we solve the tomography sub-problem approximately, by using only a few number of CG iterations since this strategy in practice greatly improves ADMM convergence rates, see review in~\cite{Boyd:11}. 

The minimization problem with respect to $\psi$ \eqref{Eq:reg} has a closed form solution defined via the soft-thresholding operator~\cite{Donoho:95},
\begin{equation}
  \tilde{\psi} = \frac{\nabla \mu^{k+1}+2\lambda^k/\rho}{|\nabla \mu^{k+1}+2\lambda^k/\rho|}\max(0,|\nabla \mu^{k+1}+2\lambda^k/\rho|-2\alpha/\rho).
  \label{Eq:solreg}
\end{equation}

\begin{figure}
    \centering
    \includegraphics[width=1\textwidth]{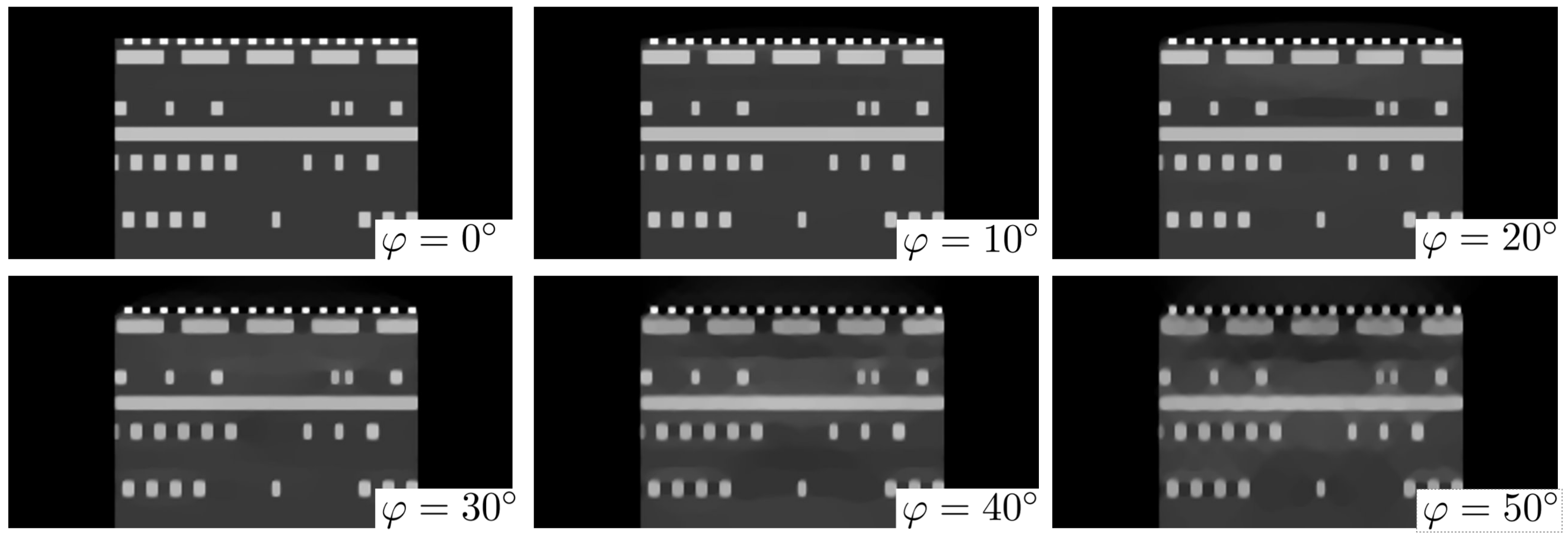}
    \caption{Results of reconstruction with TV regularization for different laminography tilt angles. Data simulation parameters are the same as in Figure~\ref{fig:undersamplng}.}
    \label{fig:recreg}
\end{figure}

Figure~\ref{fig:recreg} demonstrates results of solving the minimization problem~\eqref{eq:mainregproblem} for different laminography tilt angle $\varphi$. Data simulation parameters for the integrated circuit dataset were the same as in Figure~\ref{fig:undersamplng}. The regularization parameter $\alpha$ was chosen using the L-curve criterion~\cite{Agarwal:03}.

Comparing reconstructed vertical slices presented in this figure to the ones presented in Figure~\ref{fig:undersamplng}c, one can see that the proposed TV regularization approach significantly helps in suppressing the 'tail' artefacts even for large laminography angles. It should be noted, that the TV regularization for large values of the regularization parameter may also introduce cartoon-like artefacts~\cite{chan2005image}. These artefacts are particularly seen in reconstruction for $\varphi=50^\circ$ where we used a higher value of $\alpha$ than for other cases. It is generally advisable to search for a trade-off between the 'tail' and 'cartoon' artefacts with varying values of $\alpha$.

\end{document}